\documentclass[onecolumn,french,english,latin9,bibyear, twocolumn]{aa}
\usepackage[latin9]{inputenc}
\usepackage{textcomp}
\usepackage{amstext}
\usepackage{graphicx}
\usepackage{esint}

\makeatletter

\newcommand{\noun}[1]{\textsc{#1}}
\newcommand{\lyxmathsym}[1]{\ifmmode\begingroup\def\b@ld{bold}
  \text{\ifx\math@version\b@ld\bfseries\fi#1}\endgroup\else#1\fi}

\providecommand{\tabularnewline}{\\}

\author{F. Raucq$^{(1)}$, G. Rauw$^{(1)}$, E. Gosset$^{(1)}$, Y. Naz\'e$^{(1)}$, L. Mahy$^{(1)}$, A. Herv\'e$^{(2)}$,
and F. Martins$^{(2)}$}

\institute{$^{(1)}$Department of Astrophysics, Geophysics and Oceanography,
Li\`ege University, Li\`ege, BELGIUM}

\institute{$^{(2)}$LUPM, Montpellier University 2, Montpellier, France}
\authorrunning{F. Raucq et al.}
\titlerunning{Signatures of past mass exchange in LSS~3074}
\offprints{F.\ Raucq}
\mail{fraucq@doct.ulg.ac.be}
\keywords{Stars: early-type -- binaries: spectroscopic -- Stars: fundamental parameters -- Stars: massive -- Stars: individual: LSS~3074}

\abstract{The role of mass and momentum exchanges in close massive binaries is very important in the subsequent evolution of the components. Such exchanges produce several observational signatures such as asynchronous rotation and altered chemical compositions, that remain after the stars detach again.}
{We investigated these effects for the close O-star binary LSS~3074 (O4\,f + O6-7\,:(f):), which is a good candidate for a past Roche lobe overflow (RLOF) episode because of its very short orbital period, \it  P \rm = 2.185\,days, and the luminosity classes of both components.}
{We determined a new orbital solution for the system. We studied the photometric light curves to determine the inclination of the orbit and Roche lobe filling factors of both stars. Using phase-resolved spectroscopy, we performed the disentangling of the optical spectra of the two stars. We then analysed the reconstructed primary and secondary spectra with the CMFGEN model atmosphere code to determine stellar parameters, such as the effective temperatures and surface gravities, and to constrain the chemical composition of the components.}
{We confirm the apparent low stellar masses and radii reported in previous studies. We also find a strong overabundance in nitrogen and a strong carbon and oxygen depletion in both primary and secondary atmospheres, together with a strong enrichment in helium of the primary star.}
{We propose several possible evolutionary pathways through a RLOF process to explain the current parameters of the system. We confirm that the system is apparently in overcontact configuration and has lost a significant portion of its mass to its surroundings. We suggest that some of the discrepancies between the spectroscopic and photometric properties of LSS 3074 could stem from the impact of a strong radiation pressure of the primary.}

\usepackage{babel}

\usepackage{babel}

\usepackage{babel}

\usepackage{babel}

\makeatother

\usepackage{babel}
\addto\extrasfrench{%
   \providecommand{\fg}{\ifdim\lastskip>\z@\unskip\fi~\frqq}%
}

\begin{document}

\title{\textmd{Observational signatures of past mass-exchange episodes in
massive binaries : The case of LSS~3074}%
\thanks{Based on observations collected at the European Southern Observatory
(La Silla, Chile) and the Cerro Tololo Inter-American Observatory
(Chile). CTIO is a division of the National Optical Astronomy Observatory
(NOAO). NOAO is operated by the Association of Universities for Research
in Astronomy (AURA), Inc., under cooperative agreement with the National
Science Foundation (USA). Also based on observations collected with
XMM-Newton, an ESA science mission with instruments and contributions
directly funded by ESA member states and the USA (NASA).%
}\textmd{$^{,}$}%
\thanks{Tables A1, A2 and A3 are only available in electronic form at the
CDS via anonymous ftp to cdsarc.u-strasbg.fr (130.79.128.5) or via
http://cdsweb.u-strasbg.fr/cgi-bin/qcat?J/A+A/ %
}}

\author{F.~Raucq\inst{\ref{inst1}}, E.~Gosset\inst{\ref{inst1}},
G.~Rauw\inst{\ref{inst1}}, J.~Manfroid\inst{\ref{inst1}},
L.~Mahy\inst{\ref{inst1}}, N.~Mennekens\inst{\ref{inst2}},
and D.~Vanbeveren\inst{\ref{inst2}}}

\institute{Space sciences, Technologies and Astrophysics Research (STAR) Institute,
Li\`ege University, Quartier Agora, All\'ee du 6 Ao\^ut 19c, B\^at. B5c, B-4000
Li\`ege, Belgium \label{inst1}\and Astronomy and Astrophysics Research
Group, Vrije Universiteit Brussel, Pleinlaan 2, 1050 Brussels, Belgium
\label{inst2}}

\mail{fraucq@doct.ulg.ac.be}

\maketitle
\makeatother

\section{Introduction}

Some recent studies have shown that a high percentage of massive stars
belong to binary or higher multiplicity systems (e.g.\ Mason et al.~\cite{Mason98},
Sana et al.\ \cite{Sana}, Sana et al.~\cite{Sana14}, Sota et al.~\cite{Sota14}).
This multiplicity enables us to observationally determine the minimum
masses of the stars through their orbital motion, but it also influences
the evolution of the stars in various ways (e.g.\ Langer \cite{Langer1}).
These evolutionary effects range from tidally induced rotational mixing
(e.g.\ de Mink et al.\ \cite{deMink}), over exchange of matter
and angular momentum through a Roche lobe overflow (RLOF) interaction
(e.g.\ Podsiadlowski et al.\ \cite{Podsiadlowski}, de Loore \&
Vanbeveren \cite{dLV}, Wellstein et al.\ \cite{Wellstein}, Hurley
et al.\ \cite{Hurley}), to the merging of both stars (e.g.\ Podsiadlowski
et al.\ \cite{Podsiadlowski}, Wellstein et al.\ \cite{Wellstein}).
In RLOF interactions, one distinguishes three different situations:
case A, if the RLOF episode occurs when the mass donor is on the core
hydrogen-burning main sequence; case B, when the star is in the hydrogen
shell burning phase; and case C, when the star is in the helium shell
burning phase (Kippenhahn \& Weigert \cite{KW}, Vanbeveren et al.\ \cite{VDLVR}).
Such binary interactions significantly affect the physical properties
of the components and their subsequent evolution. Despite considerable
progresses in theoretical models, a number of open issues, such as
the actual efficiency of accretion, remain (e.g.\ Wellstein et al.\ \cite{Wellstein},
de Mink et al.\ \cite{deMink1}, Dray \& Tout \cite{DT}). To better
understand this phenomenon, in-depth studies of systems undergoing
or having undergone mass exchange are needed.

\medskip{}

In this context, the short-period spectroscopic binary LSS~3074 (also
identified as V889 Cen and ALS 3074), classified as O4f\textbf{$^{+}$}
+ O6-7:(f): (Morrell \& Niemela \cite{Morrell}), with $P_{{\rm orb}}=2.185$
days, is an extremely interesting target. This system is located behind
the Coalsack region. The distance of this region was evaluated as
188 $\pm$ 4 pc (Seidensticker \cite{Seidensticker} and Seidensticker
\& Schmidt-Kaler \cite{Seid-Schmi}). LSS~3074 harbours one of the
very few known O4f stars. These objects are rare since they probably
represent a short-lived transition phase in the evolution of massive
O-type stars before they become Wolf-Rayet stars. In addition, the
very short orbital period of this system makes it a good candidate
for a past RLOF episode. Obtaining a good orbital and photometric
solution for this system and determining the fundamental properties
of its components are therefore of the utmost importance to better
understand these transition objects.

\medskip{}

A first preliminary orbital solution of LSS 3074 was presented by
Morrell \& Niemela (\cite{Morrell}). The low $m$ sin$^{3}i$ values
inferred from this orbital solution (8 and 9 M$_{\odot}$) are rather
surprising for such early-type objects. Niemela et al.\ (\cite{Niemela})
provided an improved orbital solution and they discussed the phase-locked
polarization variability of this system. Fitting a model to these
variations, they inferred an inclination of $i=75\text{\textdegree}$,
yielding very low absolute masses of 10-11 M$_{\odot}$ and 11-12
M$_{\odot}$ for the O4f$^{+}$ and the O6-7 component, respectively.
Niemela et al.\ cautioned however that tidal deformations could introduce
additional polarization, biasing the inclination towards 90\textdegree{}.
Optical light variations of LSS 3074 were first reported by Haefner
et al.\ (\cite{Haefner}), although these authors did not achieve
a phase coverage of the light curve allowing them to clearly distinguish
ellipsoidal variations from photometric eclipses. Haefner et al.\ suggested
an inclination of 50-55\textdegree{}, yielding again rather low masses
of 17-21 M$_{\odot}$ for both components.

\medskip{}

In the present study, we discuss our determination of the orbital
solution of this system and of the fundamental parameters of its components
through several analysis techniques. Some very preliminary results
were given in Gosset et al.\ (\cite{Gosset05}). The data used in
our study are discussed in Sect.\,\ref{data}. In Sect.\,\ref{optical-spectrum}
and Sect.\,\ref{orb_sol}, we present the optical spectrum of LSS~3074
and our determination of its orbital solution. In Sect.\,\ref{Line-profile-variability},
we perform a line profile variability study of several important lines
of the optical spectrum of LSS~3074. In Sect.\,\ref{Prelim}, we
present the preparatory treatment of our data, including the disentangling
of the observed spectra to reconstruct individual spectra of the binary
components needed for the subsequent spectral analysis, and in Sect.\,\ref{photometry},
we analyse the light curve. The spectral analyses, carried out with
the non-LTE model atmosphere code CMFGEN, are presented in Sect.\,\ref{Modelatmosphere}.
In Sect.\,\ref{Conclusions}, we conclude by a discussion concerning
the evolutionary status of LSS~3074.

\section{Observations and data reduction\textmd{ \label{data}}}

\subsection{Spectroscopy}

Optical spectra of LSS 3074 were gathered with different instruments
during several observing runs between 2002 and 2004 (see Table~\ref{journal}).

\begin{table*}[tbh]
\caption{Journal of the spectroscopic observations of LSS 3074.\label{journal}}

\begin{centering}
\begin{tabular}{c||ccc||cc||cc}
\hline 
HJD-2\,450\,000  & Instrument  & Exp. time & $\phi$  & RV$_{1}$ & RV$_{2}$ & RV$_{1,{\rm corr}}$ & RV$_{2,{\rm corr}}$\tabularnewline
 &  &  (min.) &  & (km~s$^{-1}$) & (km~s$^{-1}$) & (km~s$^{-1}$) & (km~s$^{-1}$)\tabularnewline
\hline 
\hline 
2353.719 & EMMI & 60 & 0.48 & -68.9 &  & -70.3 & \tabularnewline
2354.691 & EMMI & 60 & 0.93 & -183.6 & 72.4 & -184.2 & 78.6\tabularnewline
2355.720 & EMMI & 60 & 0.40 & 108.3 & -160.7: & 101.2 & -154.4:\tabularnewline
2381.596 & FEROS & 60 & 0.24 & 165.8 & -203.6 & 161.9 & -197.4\tabularnewline
2382.597 & FEROS & 60 & 0.70 & -246.7: & 170.8: & -251.8: & 159.1:\tabularnewline
2383.598 & FEROS & 60 & 0.15 & 127.9: & -178.9: & 123.2: & -172.6:\tabularnewline
2783.576 & FEROS & 60 & 0.19 & 169.9: & -192.9 & 165.2: & -196.7\tabularnewline
2784.620 & FEROS & 60 & 0.67 & -255.6: & 177.8: & -256.1: & 173.8:\tabularnewline
3130.613 & FEROS & 60 & 0.01 & -51.3 &  & -51.8 & \tabularnewline
3131.594 & FEROS & 75 & 0.46 & -31.8 &  & -29.1 & \tabularnewline
3132.576 & FEROS & 75 & 0.90 & -198.0 & 77.3: & -203.1 & 69.1:\tabularnewline
3133.629 & FEROS & 40 & 0.39 & 81.5 & -156.3 & 76.9 & -161.3\tabularnewline
3133.660 & FEROS & 40 & 0.40 & 68.5 & -151.3: & 63.8 & -156.3:\tabularnewline
3134.578 & FEROS & 40 & 0.82 & -266.7 & 154.7 & -267.0 & 151.3\tabularnewline
3134.608 & FEROS & 40 & 0.83 & -257.8 & 162.8 & -258.3 & 160.8\tabularnewline
\end{tabular}
\par\end{centering}

\centering{}\tablefoot{The phases ($\phi$) are computed according
to the ephemerides listed in Table\,\ref{table_solorb}. The radial
velocities are presented in Sect.~\ref{orb_sol}. The typical uncertainties
on the RVs are $10-15$ km~s$^{-1}$. The colons indicate uncertainties
larger than 20 km~s$^{-1}$.}
\end{table*}

Three spectra were obtained with the EMMI instrument on the New Technology
Telescope (NTT) at the European Southern Observatory (ESO) at La Silla
during a three-day observing run in March 2002. The EMMI instrument
was used in the \'echelle mode with grating \#9 and grism \#3, providing
a resolving power of 7700. The data were reduced using the ECHELLE
context of MIDAS, and 16 usable spectral orders, covering the wavelength
domain from 4040 to 7000 \AA{}, were extracted and normalized individually.

Twelve \'echelle spectra of LSS 3074 were taken with the Fiber-fed Extended
Range Optical Spectrograph (FEROS; Kaufer et al.\ \cite{Kaufer}).
In April 2002, the spectrograph was attached to the ESO 1.52 m telescope
at La Silla, while in May 2003 and May 2004, it was used at the 2.2
m ESO/MPE telescope at La Silla. The exposure times were 40 - 75 minutes.
The detector was an EEV CCD with 2048 $\times$ 4096 pixels of $15\,\mu$m$\times15\,\mu$m.
We used an improved version of the FEROS context within the MIDAS
package provided by ESO to reduce the data (Sana et al.\ \cite{Sana06a}).

\subsection{Photometry\label{Photometry}}

During March-April-May 2001, LSS\,3074 was observed in photometry
with the Yale 1m telescope at the Cerro Tololo Interamerican Observatory.
The telescope was operated in service mode (project A01A0098, PI E.\,Gosset)
by the YALO consortium. The telescope was equipped with the imaging
camera \textit{{ANDICAM}}.

Beyond some shortening due to bad weather conditions, the run was
initially split into two periods: from HJD 2~451~992 (2001-03-24)
to HJD 2~452~003 (2001-04-04) and from HJD 2~452~023 (2001-04-24)
to HJD 2~452~037 (2001-05-08). It was designed to optimize the Fourier
spectral window for period determination with the least observing
effort, while reaching a frequency resolution corresponding to a total
duration of about 50 days without being hampered by the aliasing due
to the central gap. In order to reduce the effect of the one-day aliasing
in the photometric data, we tried to observe the star three times
per night roughly at 3 hours before meridian, at meridian and at 3
hours after it.

The \textit{{ANDICAM}} camera was equipped with a Loral CCD of 2048
by 2048 15$\mu$m pixels. The projection of a pixel on the sky corresponds
to 0.30\arcsec\ allowing a convenient sampling. The whole field
was thus covering a 10\arcmin\ by 10\arcmin\ square on the sky.
The CCD was read out with two amplifiers. At the time of the observations
it was characterized by a 11 e$^{-}$/pixel readout noise; the gain
was set to 3.6 e$^{-}$ per ADU. Half of the CCD presented a variable
noise pattern but the amplitude was sufficiently small not to be a
concern. Moreover, LSS\,3074 was systematically centred on the best
half of the CCD.

The various measurements were performed using the Johnson system $B$,
$V$, $R$, and $I$ filters. Each measurement consisted of exposures
of 2$\times$10\,s in $B$, 2$\times$3\,s in $V$, 1$\times$1\,s
in $R$, and 2$\times$1\,s in $I$. For each night, a minimum of
five dome flat fields in each filter and of 20 bias frames were acquired.
A few Stetson standard fields (PG0918, PG1047, PG1323, PG1633, and
SA107) were observed on three good nights (May 4, 5, and 8) to calibrate
the above-mentioned photometric observations.

We reduced the raw data in the standard way (bias subtraction, division
by a combined flat field, etc.) independently for the two halves of
the CCD. The list of objects were built and the relevant instrumental
magnitudes were extracted via DAOPHOT (Stetson \cite{Stetson}) under
IRAF with both an aperture integration approach and a psf-fitting
technique. A consistent natural system was established using a multi-night,
multi-star, and multi-filter method as described in Manfroid (\cite{Manfroid1};
see also Manfroid et al.\ \cite{Manfroid2}). A set of constant stars
was iteratively constructed to perform precise relative photometry.
The differential data were calculated with an aperture radius of 1.5\arcsec
. A least-square fit of the standard star data allowed us to obtain
absolute zero points and colour transformation coefficients using
the large aperture (3.3\arcsec) data. The derived colour transformation
equations were

\begin{equation} 
	\begin{split}
		(B-V)_{std}\, =\, & 0.0236\,(\pm0.0255)\\
					&+0.9991\,(\pm0.0323)\,(B-V)_{ctio}\\
	\end{split}
\end{equation} 
\begin{equation} 
	\begin{split}
		(V-R)_{std}\, =\, & 0.1016\,(\pm0.0153)\\
					&+0.7804\,(\pm0.0316)\,(V-R)_{ctio}\\
	\end{split}
\end{equation} 
\begin{equation} 
	\begin{split}
		(V-I)_{std}\, =\, & 0.0404\,(\pm0.0237)\\
					&+0.9396\,(\pm0.0249)\,(V-I)_{ctio}\\
	\end{split}
\end{equation} 
\begin{equation} 
	\begin{split}
		V_{std}\, =\, & -0.0145\,(\pm0.0181)\\
				&+-0.0219\,(\pm0.0229)\,(B-V)_{ctio}+V_{ctio}
	\end{split}
\end{equation} 

\medskip{}

The internal precision of the data corresponds to $\sigma$\,=\,0.007,
0.010, 0.013, and 0.013 mag for the $B$, $V$, $R$, and $I$ filters
respectively, whereas the accuracy of the absolute magnitudes turned
out to be $\sigma$\,=\,0.02 mag. The measurements for LSS\,3074
are available at the CDS in Table~A1. During the iterative removing
of the variable stars, two stars were rejected early in the process
and turned out to be significantly variable, i.e.\ HD\,116827 and
LSS\,3072. They are further discussed in the Appendix and the corresponding
measurements are available at the CDS in Tables~A2 and A3.

\subsection{\textit{XMM-Newton} observations \label{obsxmm}}

\textit{XMM-Newton} (\cite{Jansen}) observed LSS~3074 twice. The
first observation (ObsID 0109100201), taken in August 2001, was centred
on LSS~3074 itself. The three EPIC (\cite{pn,MOS}) cameras were
operated in full-frame mode and the medium filter was used to discard
optical and UV light. The corresponding X-ray image of the field can
be found in Gosset et al.\ \cite{Gosset05}. A second observation
(ObsID 0036140201), centred on the low-mass X-ray binary XB~1323-619,
was taken in January 2003. This time, the EPIC-pn camera and the central
chip of the EPIC-MOS1 were operated in timing mode, whilst the EPIC-MOS2
was used in full-frame mode. All EPIC cameras were used in combination
with the thin filter. The detectors used in timing mode do not provide
data concerning LSS~3074. But since the star is located relatively
far off axis in the second observation, it fell on one of the peripheral
CCD chips of both MOS detectors. The raw data were processed with
SAS v15.0.0 using calibration files available in June 2016 and following
the recommendations of the \textit{XMM-Newton} team%
\footnote{http://xmm.esac.esa.int/sas/current/documentation/threads/ %
}. We notably filtered the data to keep only best-quality data (\textsc{{pattern}}
of 0--12 for EPIC-MOS and 0--4 for EPIC-pn data). Whilst the background
level was high during the first observation, no genuine background
flares due to soft protons affected either of the two observations.
In the first observation, LSS~3074 falls very close to the out-of-time
events of the very bright X-ray binary in the MOS2 data, thus rendering
these specific data difficult to use.

We extracted the EPIC spectra of LSS~3074 via the task \textit{especget}.
For the source regions, we used a circular region with radius 30\arcsec,
centred on the Simbad coordinates of the binary. The background was
evaluated over an annulus with inner and outer radii of 30 and 45\arcsec,
respectively. Specific ARF and RMF response files were computed to
calibrate the flux and energy axes. The EPIC spectra were grouped
with the SAS command \textit{specgroup} to obtain an oversampling
factor of five and to ensure that a minimum signal-to-noise ratio
of 3 (i.e.\ a minimum of 10 counts) was reached in each spectral
bin of the background-corrected spectra.

\begin{table}
\caption{Journal of X-ray observations of LSS 3074.}

\resizebox{9cm}{!}{%
\begin{tabular}{c|cccc}
\hline 
Rev. & Instrument & Duration & JD(start) & JD(end)\tabularnewline
 &  & (ks) & -2 450 000 & -2 450 000\tabularnewline
\hline 
0309 & MOS1 \& pn & 9.9 \& 6.0 & 2138.830 & 2138.945\tabularnewline
0575 & MOS1 \& MOS2 & 50.5 \& 50.8 & 2668.865 & 2669.452\tabularnewline
\hline 
\end{tabular}}
\end{table}

The EPIC spectrum of LSS~3074 peaks between 1.0 and 2.0\,keV. It
contains very few photons at energies below 1.0\,keV (mainly as a
result of the heavy absorption) and displays very weak emission at
energies above 2\,keV. The fits of the X-ray spectra were performed
with \texttt{xspec} (\cite{Arnaud}) version 12.9.0i. To evaluate
the neutral hydrogen column density due to the interstellar medium
(ISM), we adopt $B-V=1.44$. Accounting for the intrinsic $\left(B-V\right)_{0}$
quoted by \cite{MP} and using the conversion between colour excess
and neutral hydrogen column density of \cite{Bohlin}, we estimate
$N_{H}=0.99\,10^{22}$\,cm$^{-2}$. The X-ray absorption by the ISM
was modelled using the T\"ubingen-Boulder model (\cite{Wilms}). X-ray
spectra from massive stars can be further absorbed by the material
of the ionized stellar wind. To model such an absorption, we imported
the stellar wind absorption model of \cite{HD108} into \texttt{xspec}
as a multiplicative tabular model (hereafter labelled as \texttt{wind}).
To the first approximation, the emission from massive stars can be
represented by models of collisionally ionized equilibrium optically
thin thermal plasmas. In our fits, we used \texttt{apec} models (\cite{apec})
computed with ATOMBD v2.0.2 as provided within \texttt{xspec}. The
plasma abundances were taken to be solar (\cite{Asplund}). Given
the rather low quality of the spectra, reasonable fits were obtained
for models of the kind \texttt{tbabs$\times$wind$\times$apec}. The
results are listed in Table\,\ref{specX}. Whilst the best-fit parameters
of the two observations differ, these differences could be due to
the ambiguity between a soft, highly absorbed plasma and a harder,
less absorbed plasma. We have thus also performed a fit of the combined
datasets. The results are again quoted in Table\,\ref{specX}.

We find that the X-ray flux corrected for the ISM absorption is close
to $8.6\,10^{-14}$\,erg\,cm$^{-2}$\,s$^{-1}$. If we consider
the $B$ and $V$ magnitudes of LSS~3074 along with the bolometric
corrections from \cite{MP}, we obtain $\log{L_{{\rm X}}/L_{{\rm bol}}}\simeq-7.3$
which, given the uncertainties on both the X-ray and bolometric fluxes,
is entirely compatible with the canonical value observed for most
O-type stars (Sana et al.\ \cite{Sana06b}, \cite[and references therein]{YN}).

\begin{table*}
\caption{X-ray spectral fits of LSS~3074.\label{specX}}

{\tiny }%
\begin{tabular}{c|cccccccc}
\hline 
{\tiny Rev. } & {\tiny $\log{N_{{\rm wind}}}$ } & {\tiny $kT$ } & {\tiny norm } & {\tiny $\chi_{\nu}^{2}$ } & {\tiny $f_{{\rm X}}$ } & {\tiny $f_{{\rm X}}^{{\rm un}}$} &  & \tabularnewline
 & {\tiny (cm$^{-2}$) } & {\tiny (keV) } & {\tiny (cm$^{-5}$) } &  & \multicolumn{2}{c}{{\tiny ($10^{-14}$\,erg\,cm$^{-2}$\,s$^{-1}$)}} &  & \tabularnewline
\hline 
{\tiny \vspace*{-2mm}
} &  &  &  &  &  &  &  & \tabularnewline
{\tiny 0309 } & {\tiny $21.83_{-.48}^{+.25}$ } & {\tiny $0.99_{-.33}^{+.65}$ } & {\tiny $(1.26_{-0.57}^{+1.06})\,10^{-4}$ } & {\tiny $1.55(7)$ } & {\tiny $3.6$ } & {\tiny $8.7$ } &  & \tabularnewline
{\tiny \vspace*{-2mm}
} &  &  &  &  &  &  &  & \tabularnewline
{\tiny 0575 } & {\tiny $22.07_{-.16}^{+.18}$ } & {\tiny $0.47_{-.18}^{+.17}$ } & {\tiny $(4.72_{-2.54}^{+24.08})\,10^{-4}$ } & {\tiny $1.55(14)$ } & {\tiny $2.6$ } & {\tiny $8.3$ } &  & \tabularnewline
{\tiny \vspace*{-2mm}
} &  &  &  &  &  &  &  & \tabularnewline
{\tiny comb. } & {\tiny $22.00_{-.16}^{+.15}$ } & {\tiny $0.57_{-.14}^{+.21}$ } & {\tiny $(2.61_{-1.03}^{+2.38})\,10^{-4}$ } & {\tiny $1.64(24)$ } & {\tiny $2.9$ } & {\tiny $8.6$ } &  & \tabularnewline
{\tiny \vspace*{-2mm}
} &  &  &  &  &  &  &  & \tabularnewline
\hline 
\end{tabular}{\tiny \tablefoot{The normalization of the apec models is given
as $\frac{10^{-14}\,\int\, n_{e}\, n_{H}\, dV}{d^{2}}$ where $d$
is the distance of the source (in cm), $n_{e}$ and $n_{H}$ are the
electron and hydrogen densities of the source (in cm$^{-3}$). The
error bars indicate 1$\sigma$ uncertainties.} }
\end{table*}

\section{Optical spectrum of LSS~3074 \label{optical-spectrum}}

\begin{figure*}
\begin{centering}
\resizebox{18cm}{!}{\includegraphics[scale=0.4]{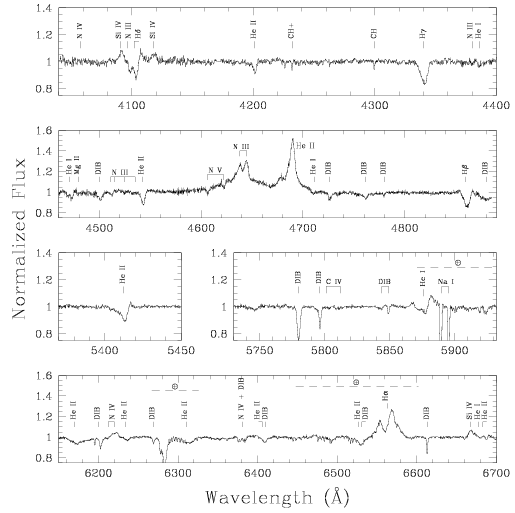}}
\par\end{centering}

\caption{Most important parts of the optical spectrum of LSS~3074 as observed
with the EMMI instrument at the NTT on HJD 2\,452\,355.7, $\phi=0.40$.\textbf{
}This spectrum is not corrected for the telluric absorption, and the
main telluric bands are shown with dashed lines (Curcio et al.\ \cite{Curcio}).\label{optspectra}}
\end{figure*}

The spectrum of LSS~3074 is illustrated in Fig.\,\ref{optspectra}.
The stellar spectrum displays many absorption lines of H\,\noun{i},
He\,\noun{i}, He\,\noun{ii}, N\,\noun{iii}, N\,\noun{iv,} and
N\,\noun{v}. The most prominent emission lines are He\,\noun{ii}
$\lambda$\,4686, N\,\noun{iii} $\lambda\lambda$\,4634-41, H$\alpha$,
and He\,\noun{i} $\lambda$\,5876. Weaker emission lines of Si\,\noun{iv}
$\lambda\lambda$\,4089, 4116, 6668, and N\,\noun{iv} $\lambda\lambda$\,4058,
6212-20 as well as P-Cygni emission components in He\,\noun{ii} $\lambda$\,5412
and H$\beta$ lines are also seen. In addition to some interstellar
absorption lines (Na\,\noun{i}, CH, CH$^{+}$), there are also numerous
(and rather strong) diffuse interstellar bands (DIBs; see Herbig \cite{Herbig}).
The strength of these features is an indication of the heavy interstellar
absorption towards LSS~3074.

The presence of N\,\noun{v} $\lambda\lambda$\,4604 and 4620 may
indicate that one of the components of LSS~3074 is rather hot (O3-O4,
Walborn \cite{Walborn}, Walborn et al.\ \cite{Walborn02}). The
radial velocities of the N\noun{\,iii} $\lambda\lambda$\,4634,
4641, and N\,\noun{iv} $\lambda$\,4058 emission lines and the N\,v
absorption lines are shown in Fig.\,\ref{RVs_N} below and analysed
in Sect.\,\ref{Spectral-types}.

Several emission lines in the spectrum of the system display strong
line profile variations. The most prominent modulations are seen in
the He\noun{\,ii} $\lambda$\,4686, He\,\noun{i} $\lambda$\,5876,
and H$\alpha$ lines (see Fig.\,\ref{figprofile}) and they are presented
in Sect.\,\ref{Line-profile-variability}.

\section{Orbital solution\textmd{\label{orb_sol}}}

Since LSS 3074 is a rather faint and heavily reddened object, the
accuracy of any radial velocity (RV) determination is mainly limited
by the S/N ratio of the data. Therefore, we have concentrated our
efforts on the strongest absorption lines that are essentially free
from blends with other features. In this way, we have measured the
RVs of H$\gamma$, He\,\textsc{i} $\lambda\lambda$\,4471, 5876,
and 7065 and He\,\textsc{ii} $\lambda\lambda$\,4542, 5412, and
6406 in the spectra. We adopted the effective wavelengths of Underhill
(\cite{Underhill}), as listed in Table~\ref{rest-wavelengths}. 

\textbf{}
\begin{table}
\textbf{\caption{Adopted wavelengths of the lines used to measure the RVs of LSS\,3074.}
}

\begin{centering}
\begin{tabular}{cc}
\hline 
Spectrum & Wavelength\,(\AA{})\tabularnewline
\hline 
\hline 
H\,\noun{i} & 4340.47\tabularnewline
He\,\noun{i} & 4471.48\tabularnewline
He\,\noun{ii} & 4541.59\tabularnewline
He\,\noun{ii} & 5411.52\tabularnewline
He\,\noun{i} & 5875.62\tabularnewline
He\,\noun{ii} & 6406.44\tabularnewline
He\,\noun{i} & 7065.24\tabularnewline
\hline 
\end{tabular}\hfill{}%
\begin{tabular}{cc}
\hline 
Spectrum & Wavelength\,(\AA{})\tabularnewline
\hline 
\hline 
N\,\noun{iv} & 4057.80\tabularnewline
N\,\noun{v} & 4603.83\tabularnewline
N\,\noun{v} & 4619.90\tabularnewline
N\,\noun{iii} & 4634.16\tabularnewline
N\,\noun{iii} & 4640.64\tabularnewline
\hline 
 & \tabularnewline
 & \tabularnewline
\end{tabular}
\par\end{centering}

\tablefoot{These effective wavelengths are taken from Underhill (\cite{Underhill}).
The top panel displays the lines we used for the determination of
the orbital solution of the system. The bottom panel shows the main
N emission and absorption lines.\label{rest-wavelengths}} 
\end{table}

For each observation, the RVs of the primary and secondary components
were computed as the mean of the corresponding RVs measured for the
above listed lines on that observation. Unfortunately, either because
of a low S/N or other problems, not all of these lines could be measured
on all of our spectra. For instance, while we managed to deblend the
primary and secondary components of the He\,\textsc{i} lines on most
of our spectra, the intensity contrast between the two components
is much larger for H$\gamma$ and for the He\,\textsc{ii} lines.
Therefore, the secondary component of the latter lines could only
be resolved on the spectra with the highest S/N ratio. Another example
is He\,\textsc{ii} $\lambda$\,5412: the proximity of several DIBs
of similar strength as the secondary line and the fact that the line
develops a P-Cygni type profile (with an emission component at a velocity
of $\sim200$\,km\,s$^{-1}$) between phase $\phi=0.35$ and $0.50$
lead to rather complex blends and thus highly uncertain RVs at these
specific orbital phases. As a result, not all of the RV data points
are based on the same set of lines. Yet, it is well known that in
early-type stars featuring relatively strong winds, the systemic velocities
of the different spectral lines can be significantly different (Rauw
et al.~\cite{HDE228766}). Therefore combining different sets of
spectral lines for different dates could bias the orbital solution.
To circumvent this problem, we tied the RVs of the different lines
to a single systemic velocity for each component. To do so, we first
noted that the He\,\textsc{i} $\lambda$\,5876 line is the one for
which RVs of both components could be determined on nearly all observations,
despite the fact that these absorptions lie on top of a broader emission.
We thus selected He\,\textsc{i} $\lambda$\,5876 as our reference
line. The positions of the primary and secondary He\,\textsc{i} $\lambda$\,5876
lines are shown in Fig.~\ref{figprofile}. For each of the other
lines that we measured, we then determined the mean RV shift compared
to He\,\textsc{i} $\lambda$\,5876, and we subtracted this mean
shift from the actual measurements. For each observation, the corrected
RVs were then averaged, resulting in a homogeneous set of values,
expressed in the rest frame of the He\,\textsc{i} $\lambda$\,5876
line.

These resulting RVs are listed in Table\,\ref{journal}. For most
data points, the uncertainties on these RVs are about 10-15 km~s$^{-1}$.
In some cases however (indicated by the colons in Table\,\ref{journal}),
the uncertainties are larger than 20 km~s$^{-1}$ and can reach $\sim$40
km~s$^{-1}$ in one case (HJD 2\,452\,382.597). To adopt the same
notations as previous investigators, we refer to the brightest and
hottest component of LSS~3074 as the primary, though its present-day
mass appears to be lower than that of its companion (see below).

\medskip{}

Owing to the severe aliasing problem, it is extremely difficult to
obtain an independent determination of the orbital period of LSS 3074
from our set of RV measurements only. A Fourier analysis of the RV$_{1}$
and RV$_{2}$ data yields the highest peaks around $\nu=0.46020$
d$^{-1}$ ($P{}_{{\rm orb}}=2.1730$ days) and $\nu=0.45762$ d$^{-1}$
($P{}_{{\rm orb}}=2.1852$ days). However, we caution that there are
many more aliases that could hide the actual orbital period. It is
worth pointing out that the second peak yields almost exactly the
same period as found by Niemela et al.\ (\cite{Niemela}). We performed
the same Fourier analysis for each of the four bandpasses of the photometric
data, $B$, $V$, $R$, and $I$, and obtained a highest peak at $\nu=0.9158$
d$^{-1}$, which corresponds to an orbital frequency of $\nu=0.9158/2=0.4579$
d$^{-1}$, corresponding to $P{}_{{\rm orb}}=2.184$ days. While the
periodogram of the photometric time series allows us to unambiguously
identify the correct alias, the natural width of its peaks is much
larger than in the case of our spectroscopic time series. In fact,
folding the RVs into the photometric period yields an unacceptably
large scatter. Therefore, combining the results from periodograms
of the photometric and spectroscopic time series, we find that the
most likely period of LSS 3074 is 2.1852 $\pm$ 0.0006 days, with
the error bars given as 1$\sigma$ deviation.

Also, the $T{}_{0}$ obtained with the photometric curves is HJD 2\foreignlanguage{french}{\emph{\,}}452\foreignlanguage{french}{\emph{\,}}000.9568,
which corresponds to a phase shift of $\sim$0.05 with the $T{}_{0}$
obtained with the RV data. If we lock the RV data to this $T{}_{0}$,
we obtain a period of 2.1849 days. Considering that this result is
well within the error bars of our previous calculations, we can admit
the phase alignment of the RV and photometric data.

As a consistency check, we also combined our measurements in the $V$
band with those taken by Haefner et al.\ in \cite{Haefner} and obtained
a highest peak of the periodogram at $\nu=0.9153$ d$^{-1}$, which
corresponds to an orbital frequency of $\nu=0.9153/2=0.45765$ d$^{-1}$,
corresponding to $P{}_{{\rm orb}}=2.1851$ days. This result tends
to confirm the quality of our determination of the orbital period
of the system.

\begin{figure}
\resizebox{9cm}{!}{\includegraphics[scale=0.4]{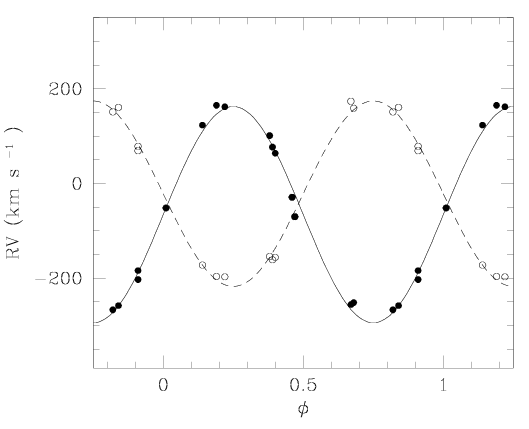}}

\caption{Radial velocities of the components of LSS 3074 assuming a period
of 2.1852 days. The RVs of the primary and secondary stars are shown
with filled and empty circles, respectively. The solid and dashed
lines indicate the orbital solution from Table\,\ref{table_solorb}.\label{fig_solorb}}
\end{figure}

Adopting an orbital period of 2.1852 days, we computed an orbital
solution assuming a circular orbit. The RVs were weighted according
to their estimated uncertainties. The result is shown in Fig.\,\ref{fig_solorb}
and the corresponding orbital elements are provided in Table\,\ref{table_solorb}.

\begin{table}
\caption{Orbital solution computed from our RV data of LSS 3074 assuming a
circular orbit and an orbital period of 2.1852 days.}

\noindent \begin{raggedright}
\begin{tabular}{l|cc}
\hline 
 & Primary & Secondary\tabularnewline
\hline 
\hline 
$T{}_{0}$ (HJD\emph{\,}-\emph{\,}2\emph{\,}450\emph{\,}000) & \multicolumn{2}{c}{2000.851 $\pm$ 0.008}\tabularnewline
$\gamma$ (km\emph{\,}s$^{-1}$) & -66.0 $\pm$ 5.0 & -21.7 $\pm$ 4.7\tabularnewline
\emph{K} (km\emph{\,}s$^{-1}$) & 228.5 $\pm$ 7.1 & 196.0 $\pm$ 6.1\tabularnewline
\emph{a\,}sin\,\emph{i} (R$_{\odot}$) & 9.9 $\pm$ 0.3 & 8.5 $\pm$ 0.3\tabularnewline
\textit{q} = \emph{m}$_{1}$/\emph{m}$_{2}$ & \multicolumn{2}{c}{0.86 $\pm$ 0.04}\tabularnewline
\emph{m}\,sin\emph{$^{3}$i }(M$_{\odot}$) & 8.0 $\pm$ 0.5 & 9.3 $\pm$ 0.7\tabularnewline
$R{}_{{\rm {RL}}}$/($a{}_{1}+a_{2}$) & 0.37 $\pm$ 0.01 & 0.39 $\pm$ 0.01\tabularnewline
$R{}_{{\rm {RL}}}$\,sin\,\emph{i }(R$_{\odot}$) & 6.7 $\pm$ 0.2 & 7.2$\pm$ 0.2\tabularnewline
$\sigma_{{\rm fit}}$ & \multicolumn{2}{c}{3.11}\tabularnewline
\hline 
\end{tabular}
\par\end{raggedright}

\tablefoot{$T{}_{0}$ refers to the time of conjunction with the
primary being in front. $\gamma$, $K$ and $a$~sin$i$ denote respectively
the apparent systemic velocity, the semi-amplitude of the radial velocity
curve and the projected separation between the centre of the star
and the centre of mass of the binary system. $R{}_{{\rm {RL}}}$ stands
for the radius of a sphere with a volume equal to that of the Roche
lobe computed according to the formula of Eggleton (\cite{Eggleton}).
All error bars indicate 1$\sigma$ uncertainties.\label{table_solorb}} 

\end{table}

The RV amplitudes of our solution are in reasonable agreement with
values from the literature ($K{}_{1}=222$ and $K{}_{2}=218$ km~s$^{-1}$;
Niemela et al.\ \cite{Niemela}). Consequently our orbital solution
also leads to low minimum masses, similar to those of Niemela et al.\ ($m$\foreignlanguage{french}{\,}sin$^{3}i=9.5$
and 10 M$_{\odot}$ for the O4f primary and the secondary, respectively).
The apparent systemic velocity of the primary is significantly more
negative than that of the secondary. This most likely indicates that
the absorption lines of the primary are not entirely formed in the
static photosphere, but arise at least partly from an expanding stellar
wind.

\begin{figure}
\resizebox{9cm}{!}{\includegraphics[scale=0.4]{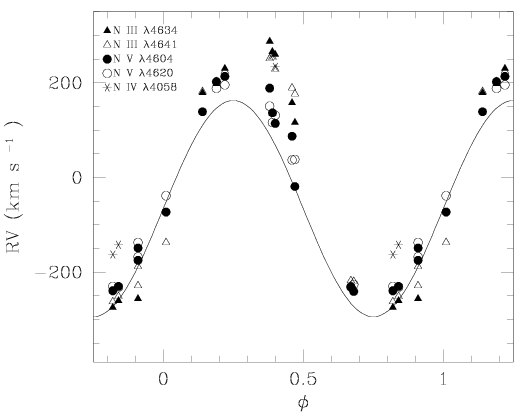}}

\caption{Radial velocities of the N\,\noun{iii} and N\,\noun{iv} emission
lines and the N\,\noun{v} absorptions in the spectrum of LSS~3074.
The meaning of the various symbols is given in the upper left corner.
The solid line yields the orbital solution of the primary from Table\,\ref{table_solorb}.\label{RVs_N}}
\end{figure}

Figure\,\ref{RVs_N} shows the RVs of the nitrogen emission and absorption
lines in the spectrum of LSS~3074. The adopted effective wavelengths
of the studied lines are listed in Table~\ref{rest-wavelengths}.
All the lines seem to move with the primary, but the scatter of the
velocities of the various lines is rather large, especially at phases
between the 0.25 quadrature and 0.5 conjunction. During this phase
interval, the N\,\noun{iii} emission lines yield systematically larger
velocities than the N\,\noun{v} absorption.

\section{Line profile variability\label{Line-profile-variability}}

\begin{figure*}
\begin{centering}
\resizebox{18cm}{!}{\includegraphics[scale=0.4]{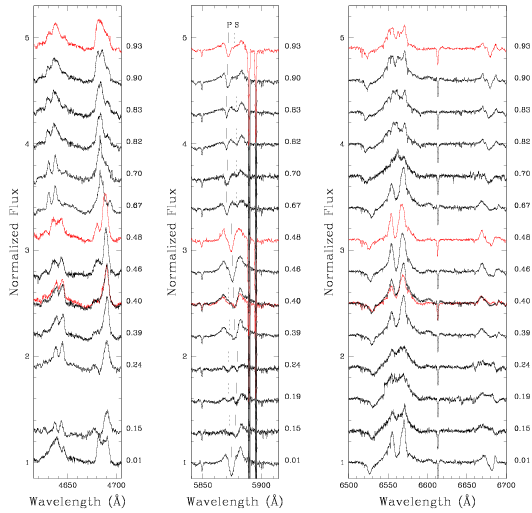}}
\par\end{centering}

\caption{Line profile variations of some important emission lines in the optical
spectrum of LSS~3074. N\,\noun{iii} $\lambda\lambda$ 4634-4641
and He\,\noun{ii} $\lambda$ 4686 are shown in the left panel; He\,\noun{i}
$\lambda$ 5876 is represented in the central panel; and H$\alpha$
is indicated in the right panel. The orbital phases computed according
to the ephemerides in Table\,\ref{orb_sol} are given on the right
of each panel. The red spectra correspond to the EMMI observations.
The vertical solid (resp.\ dotted) lines in the middle panel represent
the position of the primary (resp.\ secondary) He~\noun{i} $\lambda$
5876 line for the given observation.\label{figprofile}}
\end{figure*}

The most prominent emission lines in the optical spectrum of LSS~3074
show strong profile variations (see Fig.\,\ref{figprofile}). For
instance, the He\,\textsc{ii} $\lambda$\,4686 line evolves from
a broad and skewed emission (around $\phi=0.0$) into a double-peaked
feature with rather narrow individual peaks and the strongest peak
closely following the orbital motion of the primary (at phases near
0.5).

On the other hand, the H$\alpha$ emission features a double-peaked
profile on most of our spectra and the profiles observed around the
two conjunctions are very similar. At phases near quadrature, the
peaks are broader and less clear cut. Apart from the moving absorption
lines, the He\,\textsc{i} $\lambda$\,5876 profile has a morphology
relatively similar to that of the H$\alpha$ line.

We measured the equivalent widths (EWs) of the H$\alpha$ line between
6513 and 6604 \AA. We find that $\overline{{\rm EW}}=4.09\pm0.60$
\AA. Although the dispersion around the mean is quite large, no obvious
phase-locked behaviour is apparent. In particular, we do not see any
variations attributable to the modulation of the continuum.

\begin{figure*}[htb]
\begin{minipage}[c]{8.5cm}%
\begin{center}
\resizebox{8.5cm}{!}{\includegraphics{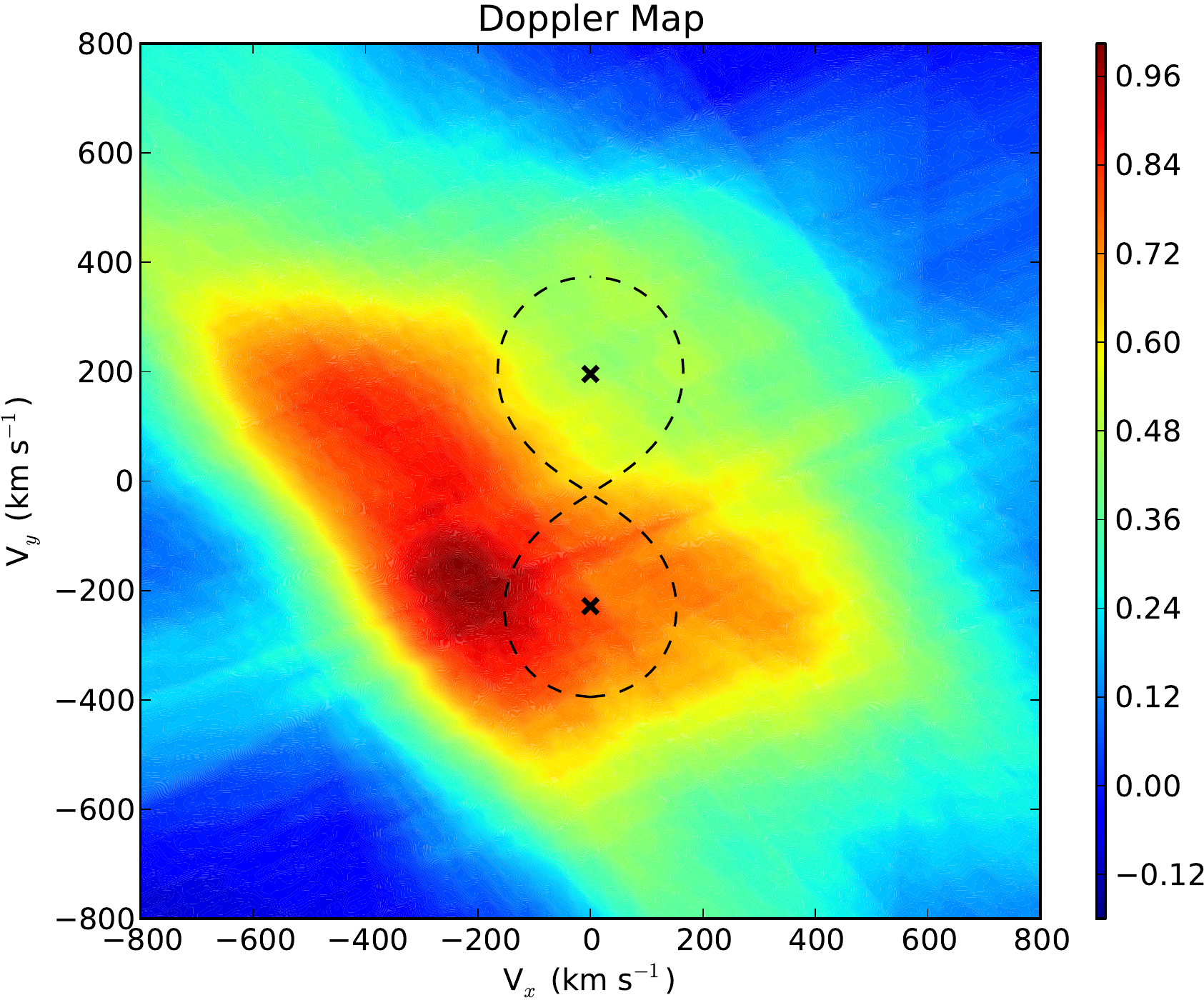}} 
\par\end{center}%
\end{minipage}\hfill{}%
\begin{minipage}[c]{8.5cm}%
\begin{center}
\resizebox{8.5cm}{!}{\includegraphics{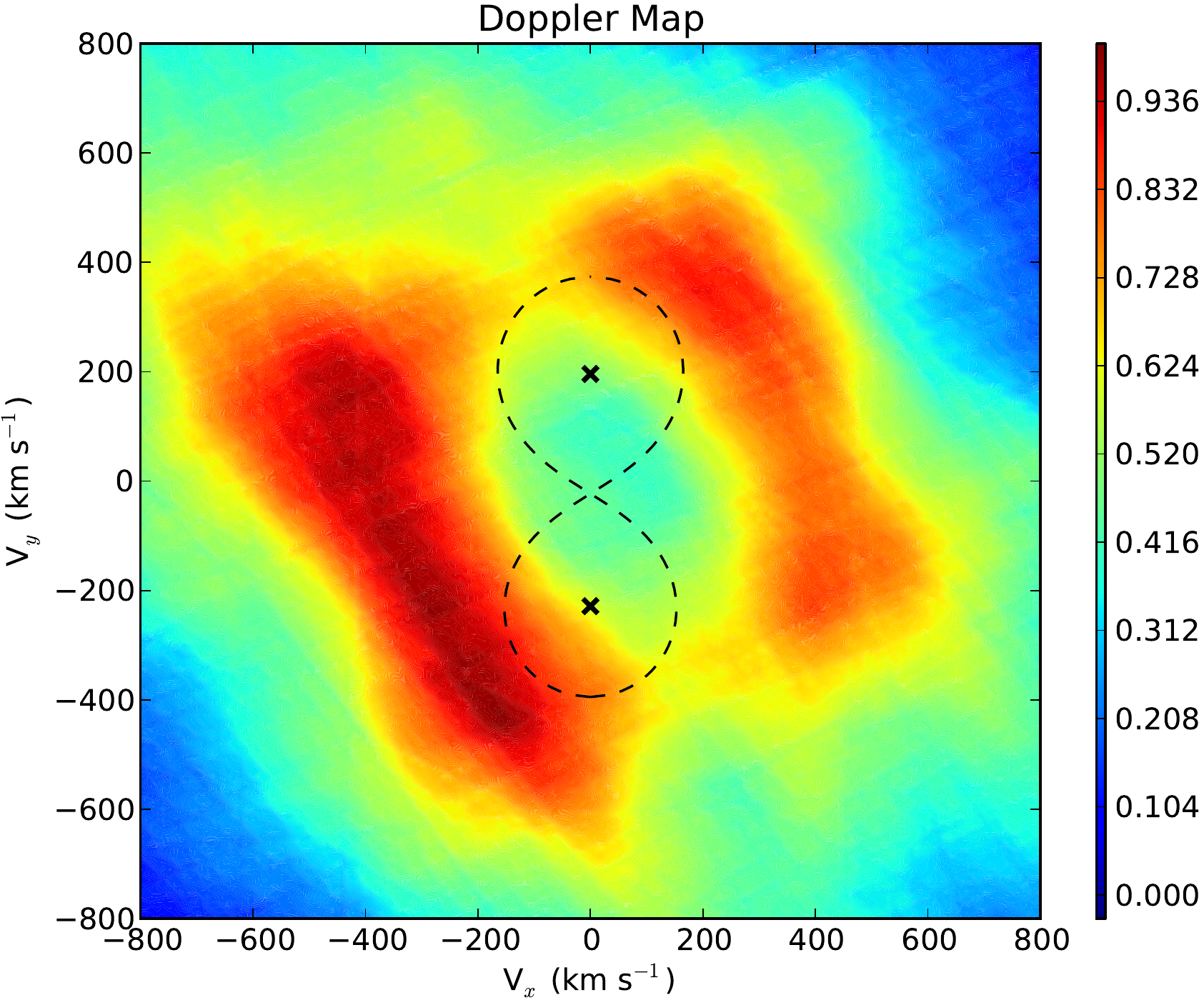}} 
\par\end{center}%
\end{minipage}\caption{Doppler maps of the He\,\textsc{ii} $\lambda$\,4686 (left) and
H$\alpha$ (right) emission lines in the spectrum of LSS~3074. The
crosses correspond to the radial velocity amplitudes of the centre
of mass of the primary (negative $v{}_{y}$) and secondary (positive
$v_{y}$), whereas the dashed lines indicate the Roche lobe in velocity
space for a mass ratio $m_{1}/m_{2}=0.86$. The colour scale indicates
the level of the line emissivity relative to its maximum value. \label{Dopplermap}}
\end{figure*}

These complex line morphologies make it impossible to assign a single
RV to the He\,\textsc{ii} $\lambda$\,4686 and H$\alpha$ emission
lines, and suggest that these lines do not arise in the atmosphere
of one of the two stars, but stem from an extended emission region.
To further quantify this situation, we used the method of Doppler
tomography to map the emitting regions of the He\,\textsc{ii} $\lambda$\,4686
and H$\alpha$ lines in velocity space. Our method is based on the
Fourier-filtered back projection technique (Horne \cite{Horne91},
Rauw et al.\ \cite{HDE228766}, Mahy \cite{Mahy11}). The radial
velocity of any gas flow that is stationary in the rotating frame
of reference of the binary can be expressed as 
\begin{equation}
v(\phi)=v_{x}\,\cos{(2\,\pi\,\phi)}-v_{y}\,\sin{(2\,\pi\,\phi)}+v_{z}\label{eqtomo}
\end{equation}
where $\phi$ stands for the orbital phase with $\phi=0.0$, corresponding
to the primary star being in front. The $(v_{x},v_{y},v_{z})$ are
the velocity coordinates of the gas flow. The $x$ and $y$ axes are
located in the orbital plane. The $x$-axis runs from the primary
to the secondary, whilst the positive $y$-axis points in the direction
of the orbital motion of the secondary. The $v_{z}$ component represents
the apparent systemic velocity of the line under consideration. The
Doppler map consists of a projection of relation given by (\ref{eqtomo})
on the $(v_{x},v_{y})$ plane. For a given value of $v_{z}$, each
pixel in a Doppler map, specified by its velocity coordinates, is
therefore associated with a particular form of equation (\ref{eqtomo}).
All our spectra listed in Table\,\ref{journal} were given equal
weights. 
The Doppler maps, computed adopting $v_{z}=-44$\,km\,s$^{-1}$,
are shown in Fig.\,\ref{Dopplermap}. They reveal extended line formation
regions in velocity space with different morphologies for the two
lines. The He\,\textsc{ii} $\lambda$\,4686 line features an emission
lobe on the negative $v_{x}$ side and mostly with negative $v_{y}$.
This emission seems thus mostly associated with the primary star.
The H$\alpha$ Doppler map displays two lobes, extending on both sides
of the Roche lobes in velocity space, with the strongest one closely
matching the emission lobe of the Doppler map of He\,\textsc{ii}
$\lambda$\,4686.

A full interpretation of these Doppler maps in terms of wind-wind
interactions is beyond the scope of the present paper. However, it
is interesting to compare our Doppler maps with those of Algol-type
systems where a cool star fills up its Roche lobe and transfers matter
to a hotter companion (Richards et al.~\cite{Richards14}). Our maps
are clearly different from those presented by Richards et al.\ (\cite{Richards14}).
The structures seen in Fig.\,\ref{Dopplermap} neither resemble those
produced by gas streams that directly impact the mass gainer nor do
they feature the ring-like structure, centred on the mass gainer,
that appears for systems with Keplerian accretion disks. These results
therefore do not support the presence of an accretion disk or an accretion
stream in LSS~3074.

\section{Preparatory analysis\textmd{ \label{Prelim}}}

\subsection{Spectral disentangling\label{disent}}

The determination of the orbital solution allowed us to recover the
individual spectra of both components through the disentangling of
the normalized spectra of the binary system. For this purpose, we
used our disentangling routine (Rauw \cite{DSc}) based on the method
of Gonz\'alez \& Levato (\cite{GL}), previously used by Linder et al.\ (\cite{Linder})
and improved by Mahy et al.\ (\cite{Mahy}). In this procedure, the
mean spectra of each binary component are reconstructed in an iterative
way by shifting the observed data into the frame of reference of one
star and subtracting the best approximation of the spectrum of the
other star shifted to its observed radial velocity. In the disentangling
carried out in this work, we fixed the RVs of the binary components
to those corresponding to the orbital solution of Sect.\,\ref{orb_sol}.
For a more detailed description of the method, see Raucq et al.\ (\cite{Raucq}).

\medskip{}

As for any disentangling method, this technique also has its limitations
(Gonz\'alez \& Levato \cite{GL}). An important limitation for our study
is that broad spectral features are not recovered with the same accuracy
as narrow ones. Indeed, features that are wider than a few times the
RV amplitude are barely properly recovered. This is the case of the
wings of the Balmer lines, for example. Moreover small residual errors
in the normalization of the input spectra can lead to oscillations
of the continuum in the resulting disentangled spectra on wavelength
scales of several dozen \AA{}, and a reasonable observational sampling
of the orbital cycle is needed because the quality of the results
depends on the radial velocity ranges covered by the observations.
Finally, spectral disentangling works on continuum-normalized spectra
and does not yield the brightness ratio of the stars, which must be
determined by other techniques (see below).

\medskip{}

In the particular case of LSS 3074, we also encountered problems due
to the large brightness ratio of the system. Indeed, as shown in Sect.\,\ref{brightness},
the primary star appears much brighter than the secondary star, meaning
that the latter is faint in the observed spectra, which induces difficulties
in the disentangling procedure, a lower S/N ratio, and thus larger
uncertainties on the resulting secondary spectrum.

\subsection{Spectral types \label{Spectral-types}}

We then determined the spectral types of the stars through the measurement
of the equivalent width ratio of some spectral lines of the reconstructed
individual line spectra of the primary and secondary components. We
used the He\,\textsc{i} $\lambda$\,4471 and He\,\textsc{ii} $\lambda$\,4542
lines, on the one hand, and Si\,\textsc{iv} $\lambda$\,4089 and
He\,\textsc{i} $\lambda$\,4143 lines, on the other hand, and applied
the Conti quantitative classification criteria for O-type stars (Conti
\& Alschuler \cite{CA}, Conti \& Frost \cite{CF}, see also van der
Hucht \cite{vdHucht}) for both spectral types and luminosity classes.
In this way, we obtain O5.5\,I and O6.5-7\,I classifications for
the primary and secondary, respectively. The strong He\,\textsc{ii}
$\lambda$\,4686 and N\,\textsc{iii} $\lambda\lambda$ 4634-41 emission
lines lead to the addition of an f qualifier (see Walborn et al.\ \cite{Walborn02},
Sota et al.\ \cite{Sota11}, Ma\'iz Apell\'aniz et al.\ \cite{Jesus}
and references therein) for both components. The primary spectrum
further displays emissions of Si\,\textsc{iv} $\lambda\lambda$ 4089,
4116. Previously these features were indicated by an f$^{+}$ qualifier,
but it has been suggested that this notation is obsolete (Sota et
al.\ \cite{Sota11}).

However, the Conti criterion for the luminosity classes is formulated
for spectral types from O7 to O9.7 and its application to the spectrum
of the primary star is thus an approximation. Moreover, Walborn et
al.\ (\cite{Walborn02}) argued that it would be preferable to use
the N\,\textsc{iv}/N\,\textsc{iii} emission line ratio for the spectral
classification of the earliest O-type stars rather than the ratio
of the He\,\textsc{i}/He\,\textsc{ii} absorption lines. As shown
in Fig.\,\ref{RVs_N}, the N\,\textsc{iii} $\lambda\lambda$ 4634,
4641, and the very weak N\,\textsc{iv} $\lambda$ 4058 emission line
(when present) clearly move along with the primary star. The same
holds true for the weak, but definite N\,\textsc{v} $\lambda\lambda$\,4604
and 4620 absorption lines. As indicated in Sect.\,\ref{optical-spectrum},
the presence of these N\,\textsc{v} absorption lines points towards
a rather hot star with an O3 to O4 spectral type (Walborn et al.\ \cite{Walborn02},
Sota et al.\ \cite{Sota11}), which is at odds with our above classification
based on the helium lines. We have thus compared our disentangled
primary spectrum with the spectral atlas of Sota et al.\ (\cite{Sota11}).
The strong He\,\textsc{ii} $\lambda$\,4686 emission clearly confirms
a supergiant luminosity class. The simultaneous presence of strong
N\,\textsc{iii}, very weak N\,\textsc{iv,} and definite N\,\textsc{v}
lines in the spectrum of the primary is clearly a challenge both for
spectral classification and spectral modelling (see Sect.\,\ref{Results}).
The best match, although certainly not perfect, is found with the
spectra of HD~14\,947 (O4.5\,If) and, to a lesser extent, HD~15\,570
(O4\,If). For a highly deformed star, such as the primary of LSS~3074,
gravity darkening leads to a non-uniform surface temperature. Hence,
the discrepancy between the O5.5\,If spectral type inferred from
the relative strengths of the He\,\textsc{i} and He\,\textsc{ii}
lines and the O4-4.5\,If type derived from the nitrogen spectrum
could simply reflect the fact that these lines form over different
parts of the stellar surface. We come back to this point in Sect.\,\ref{Results}.
For the secondary star, comparison with the atlas of Sota el al.\ (\cite{Sota11})
yields a spectral classification O6.5-7\,If, in agreement with the
results from the relative strengths of the helium lines.

Finally, since our classification is based on the disentangled spectra,
it is less sensitive to a possible phase dependence of the line strengths
and should thus be more robust than a classification based only on
spectra collected near quadrature phases.

\subsection{Brightness ratio \label{brightness}}

The spectral disentangling yields the strength of the lines in the
primary and secondary stars relative to the combined continuum. But
as we mentioned earlier, it does not allow us to establish the relative
strengths of the continua. We therefore needed to first establish
the brightness ratio of the stars to further analyse the reconstructed
spectra.

To estimate the optical brightness ratio of the components of LSS~3074,
we measured the equivalent widths, referring to the combined continuum
of the two stars, of a number of spectral lines on the reconstructed
primary and secondary spectra. The results are listed in Table\,\ref{EW}
along with the mean equivalent widths of the same lines in synthetic
spectra of stars of the same spectral type, computed with the non-LTE
model atmosphere code CMFGEN (Hillier \& Miller \cite{HM}), which
we describe in Sect.\,\ref{CMFGENcode}. For comparison, we also
list the mean equivalent widths of the same lines in spectra of stars
of similar spectral type as measured by Conti (\cite{Conti1,Conti2})
and Conti \& Alschuler (\cite{CA}).

\begin{table*}[thb]
\caption{Brightness ratio determination from the dilution of prominent lines.}

\begin{centering}
\begin{tabular}{l|ccccccc}
\hline 
Line  & \multicolumn{6}{c}{Equivalent Width (\AA{})} & $l{}_{1}/l_{2}$\tabularnewline
\hline 
 & \multicolumn{2}{c}{Observations} & \multicolumn{2}{c}{Synthetic spectra} & \multicolumn{2}{c}{Conti (\cite{Conti1,Conti2}) } & \tabularnewline
\hline 
 & Primary  & Secondary  & O5.5  & O7  & O5.5  & O7 & \tabularnewline
\hline 
\hline 
He\,\textsc{i} $\lambda$\,4026  & 0.42 & 0.19 & 0.51 & 0.69 & 0.46  & 0.61  & 3.00\tabularnewline
He\,\textsc{ii} $\lambda$\,4200  & 0.41 & 0.17 & 0.53  & 0.45 & 0.58 & 0.59 & 2.00\tabularnewline
H$\gamma$  & 0.90 & 0.34 & 1.56 & 1.75 & 1.85 & 2.22 & 3.00\tabularnewline
He\,\textsc{i} $\lambda$\,4471  & 0.25 & 0.26 & 0.29 & 0.61 & 0.29 & 0.59 & 2.08\tabularnewline
He\,\textsc{ii} $\lambda$\,4542  & 0.56 & 0.19 & 0.62 & 0.53 & 0.75 & 0.67 & 2.43\tabularnewline
\hline 
\end{tabular}
\par\end{centering}

\tablefoot{The measured EWs are compared with values for the same
lines in synthetic spectra of the same spectral type and in the compilation
of measurements from the literature. The last column yields the brightness
ratio for each line considered, using the synthetic spectra EWs. \label{EW}} 
\end{table*}

The brightness ratio of the two stars can then be evaluated from

\begin{center}
$\frac{l_{1}}{l_{2}}=(\frac{EW_{1}}{EW_{2}})_{{\rm obs}}(\frac{EW_{O7}}{EW_{O5.5}})_{{\rm mean}}$. 
\par\end{center}

By combining our measurements with those from synthetic spectra, we
derive an optical brightness ratio of $2.50\pm0.43$. As a consistency
check, we can also determine the brightness ratio through a comparison
with the measurements made by Conti, and we obtain $2.61\pm0.41$.

The disentangled continuum normalized primary and secondary optical
spectra are shown in Fig.\,\ref{fig1}.

\section{Light curve\textmd{\label{photometry}}}

To improve our understanding of the system LSS~3074, and in particular
its geometry, we performed an analysis of the photometric light curves
by modelling the system using the eclipsing binary star simulator
NIGHTFALL%
\footnote{For more details, see the Nightfall UserManual by Wichmann (1998)
available at the URL: http://www.hs.uni-hamburg.de/DE/Ins/Per/Wichmann/Nightfall.html%
}, developed by Wichmann, Kuster and Risse. We worked in an iterative
way. First of all, we fixed the mass ratio and orbital period to the
values obtained from the orbital solution (see Table\,\ref{table_solorb})
and the effective temperatures to typical values for stars of similar
spectral types (Martins et al.\ \cite{Martins2}). This permitted
us to obtain a first estimation of the photometric solution and of
the stellar radii and masses. Based on these approximated parameters,
together with the effective temperatures, we calculated the associated
surface gravities and stellar luminosities. This first approximation
of the stellar fundamental parameters was then used as starting point
input in our study of the atmosphere modelling procedure (see Sect.
\ref{CMFGENcode} and \ref{Method}). This procedure then permitted
us to accurately determine the effective temperatures of the stars,
giving us a new input for the light curve study with NIGHTFALL.

Inspection of the light curve reveals a lack of a plateau between
the minima, indicating important contributions of ellipsoidal variations
to the photometric variability. Actually, the shape of the light curve
suggests that almost the entire photometric variations could stem
from such ellipsoidal variations, thereby implying that at least one
of the stars must fill or overflow its Roche lobe. 

The only evidence for the existence of genuine eclipses comes from
the fact that the secondary minimum seems about 0.02 mag deeper than
the primary minimum. The primary minimum corresponds to the occultation
of the primary star by the secondary. What is surprising here is that
the deeper minimum actually corresponds to the spectroscopically hotter
star being in front. Given the level of dispersion in the light curve,
it is quite possible that the small difference in depth of the minima
is not significant and could arise, for example, from intrinsic variations
of the stars. An alternative explanation for this situation could
be an atmospheric eclipse produced by the wind of the primary star.
Antokhina et al.\ (\cite{Antokhina13}) studied the impact of free
electron scattering in the wind of one star on the light curves of
close eclipsing binaries. For a contact binary made of otherwise equal
stars, these authors show that the presence of a stellar wind with
$\dot{M}=10^{-5}$\,M$_{\odot}$\,yr$^{-1}$ around one component
deepens the eclipse corresponding to the star with the wind passing
in front of its companion by about 0.08\,mag. At the same time, the
depth of the other eclipse is slightly reduced. In the case of LSS~3074,
the fact that most emission lines closely follow the motion of the
primary star suggests that this star has the strongest wind, thus
lending support to this explanation. However, as pointed out by Antokhina
et al.\ (\cite{Antokhina13}), a quantitative assessment of the impact
on the light curve requires some independent determination of the
wind parameters (mass-loss rate, asymptotic velocity and exponent
of velocity law). In our case, this is difficult to achieve (see Sect.\,\ref{Modelatmosphere}).
This is because of the lack of UV spectroscopy and the complexity
of the line emission regions. Indeed, as revealed by our tomographic
analysis, at least parts of the H$\alpha$ and He\textsc{\,ii} $\lambda$\,4686
emissions arise from an interaction zone and these lines hence do
not necessarily reflect the genuine properties of the winds. Yet,
our best-fit model atmosphere parameters (see the forthcoming Table\,\ref{CMFGENparam})
yield upper limits on the wind density that could be consistent with
the observed difference in eclipse depth. However, these best-fit
parameters also suggest that both binary components lose material
at a similar rate. Hence, one would expect both eclipses to be affected
by stellar wind absorption. We thus conclude that a better knowledge
of the wind parameters of LSS~3074 is required before we can attempt
a model of the light curve accounting for the effects of the stellar
winds. Whatever the origin of the slight difference in depth of the
minima, we need to keep in mind that this situation affects the quality
of the fits of the light curve.

To explain the shape of the secondary minimum by the sole effect of
ellipsoidal variations, we at least need to assume that either both
components of the binary system fill up (contact-contact) or even
overflow their Roche lobes (overcontact), or that one of the components
must be much smaller and much fainter than the other star that fills
up its Roche lobe. In the former case, one expects both stars to be
of very comparable size and, given their effective temperatures, also
of rather comparable brightness. To explain the depth of almost 0.2
mag by the sole effect of ellipsoidal variations is actually not possible
here without requiring the presence of true eclipses. In this context,
the necessity to invoke a contact-contact or an overcontact configuration
remains valid, along with the conclusion that sizes and brightness
of both stars should be comparable.

\begin{figure}
\resizebox{9cm}{!}{\includegraphics[clip]{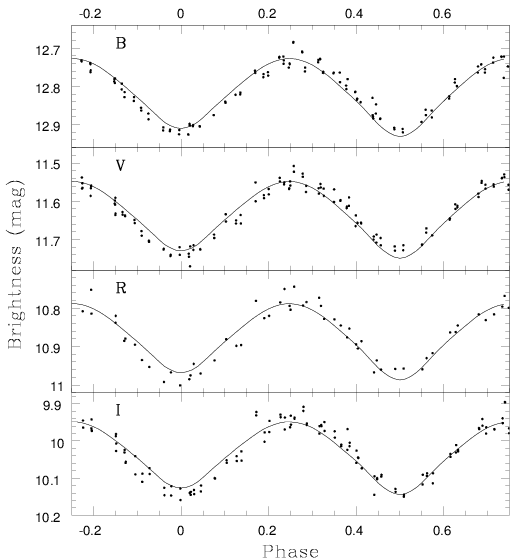}}

\caption{Photometry of LSS~3074. The observational data are presented as black
dots, and the solid black line corresponds to the best-fit theoretical
light curve fit with NIGHTFALL with the parameters presented in Table\,\ref{tab_nightfall}.
The zero phase corresponds to the secondary eclipse, when the primary
is occulting the secondary.}
\label{lightcurve}
\end{figure}

This is precisely the outcome of our fits with the NIGHTFALL code.
The best-fit quality ($\chi^2_{\nu} = 4.4$) is achieved for overcontact
configurations with a filling factor of 1.008 ($i \simeq 54.5^{\circ}$),
where the corresponding $V$-band brightness ratio is predicted to
be 1.09. The corresponding photometric solution is presented in Table~\ref{tab_nightfall}
and the associated plot is shown in Fig.~\ref{lightcurve}. Yet,
our spectroscopic analysis suggests an optical brightness ratio (primary/secondary)
near 2.50. This situation is thus clearly at odds with explaining
the photometric light curve via a double contact or overcontact configuration.
To achieve a brightness ratio of $2.50 \pm 0.43$, the ratio between
the mean radii of the primary and secondary stars would have to be
$1.15 \pm 0.10$. This translates into a ratio of the filling factors
of $1.23 \pm 0.11$, which is only possible for detached or semi-detached
configurations. As could be expected, the fact that the hotter star
is in front during the deepest minimum leads to difficulties to fit
this minimum. 

Since the eclipses are only partial, one way to solve this issue would
be to postulate the existence of a dark spot on the side of the secondary
star facing the primary. Such a spot would then lead to a deeper minimum
when the primary star is in front. From a purely numerical point of
view this would improve the fit quality significantly with $\chi^2_{\nu}$
now approaching 2.9. In addition, it would certainly help to bring
the spectroscopic and photometric brightness ratios into better agreement.
Yet, aside from the difficulty of explaining the physical origin of
such a spot, there is another issue that concerns the fact that the
spot size and its dim factor (i.e.\ the reduction in local temperature)
are not fully independent and we lack any objective constraints on
these parameters. Test calculations have shown that for some situations,
the best-fit model would actually correspond to a spot that covers
more than 75 $\%$ of the surface of the secondary, which is clearly
not physical.

\begin{table}
\caption{Photometric solution for LSS3074.}

\begin{tabular}{c|cc}
Parameters & Primary & Secondary\tabularnewline
\hline 
\hline 
\textit{i} (\textdegree{}) & \multicolumn{2}{c}{$54.5\pm1.0$}\tabularnewline
\selectlanguage{french}%
\textit{q} = \emph{m}$_{1}$/\emph{m}$_{2}$\selectlanguage{english}%
 & \multicolumn{2}{c}{\selectlanguage{french}%
0.86 (fixed)\selectlanguage{english}%
}\tabularnewline
Filling factor%
\footnote{The filling factor of a given binary component is defined here as
the fraction of its polar radius over the polar radius of its Roche
lobe.%
} & $1.008\pm0.010$ & $1.008\pm0.010$\tabularnewline
$T_{\rm eff}$\,(K) & 39~900 (fixed) & 34~100 (fixed)\tabularnewline
$m$\,(M$_{\odot}$) & $14.8\pm1.1$ & $17.2\pm1.4$\tabularnewline
$R_{\rm pole}$\,(R$_{\odot}$) & 7.8 & 8.4\tabularnewline
$\chi^{2}$ & \multicolumn{2}{c}{1820.7}\tabularnewline
$N_{{\normalcolor {\rm {dof}}}}$ & \multicolumn{2}{c}{415}\tabularnewline
\hline 
\end{tabular}

\tablefoot{$N_{{\normalcolor {\rm {dof}}}}$ is the number of degrees
of freedom.}

\label{tab_nightfall}
\end{table}

As an alternative to the overcontact configuration, one might consider
a scenario where the primary star overflows its Roche lobe and transfers
matter to a geometrically thick accretion disk around the more massive
secondary star. This kind of scenario has been proposed to explain
the light curves of $\beta$~Lyrae (Wilson \cite{Wilson}), RY~Sct
(Antokhina \& Kumsiashvili \cite{RYSct}), and V455\,Cyg (Djura\v{s}evi\'{c}
et al.\ \cite{V455Cyg}), among others. On the positive side of such
a scenario, it might help reconcile the photometric and spectroscopic
brightness ratios. On the negative side, we have the lack of a clear
signature of an accretion disk in the Doppler maps of the H$\alpha$
line and the fact that LSS~3074 is a very compact system, leaving
very little room for the formation of a disk, unless the secondary
star is very small. To test this scenario, we used the disk option
in the NIGHTFALL code. In addition to the effective temperatures that
we fixed at the values obtained from our model atmosphere fits (see
Sect.\,\ref{Method}), we set the Roche lobe filling factor of the
primary star to unity and that of the secondary star to 0.75. This
choice results in a brightness ratio that is consistent with our spectroscopic
value. We tested various options for the disk available in the NIGHTFALL
code (simple disk with uniform temperature, isothermal disk and reprocessing
disk), which correspond to different prescriptions of the variations
of the disk height and temperature with radius. Since the disk in
our models resembles more an annulus than a genuine disk, these three
options actually yield nearly identical results. These models fail
to reproduce the width of the secondary minimum. This failure increases
the $\chi^{2}$ of the fits. Still, we managed to find fits with $\chi_{\nu}^{2}\sim4.6$,
but this was only possible for a disk temperature exceeding the temperatures
of both stars, which is most probably not physical%
\footnote{Raising the disk temperature allows the model to better reproduce
the depth of the primary minimum, hence partially compensating for
the increase of $\chi^{2}$ due to the poor fit of the secondary minimum.%
}. Better quality fits can be obtained if the Roche lobe filling factor
of the secondary is allowed to drop below 0.75: we obtained $\chi_{\nu}^{2}=3.9$
for a secondary filling factor of 0.47 (which would correspond to
a secondary radius of only 3.9\,R$_{\odot}$) and a disk temperature
of 51\,300\,K. Both parameters (secondary radius and disk temperature)
are most probably not physical. We thus conclude that a disk model
cannot solve the issues raised above any better than the pure Roche
potential model.

From a purely phenomenological point of view, we can expand the light
curve into a sum of sine functions. We tried a period of 2.1852/2
= 1.0926\,days and a mere sine and a period of 2.1852\,days and
the fundamental sine plus its harmonic. The results are given in Table
\ref{tab_sinus}. The photometric observations exhibit a variation
with a semi-amplitude of 0.09\,mag in all filters. The semi-amplitude
associated with the modulation on the orbital period is 0.01\,mag
reflecting the above-mentioned difference between the depths of the
two minima. The corresponding $\chi_{\nu}^{2}$ are lower than for
any other fit we tried but still not satisfactory. Including a larger
number of harmonics in our expansion does not improve the situation.
However, in terms of standard binary models, this fit is unphysical
since a modulation with a semi-amplitude as large as 0.09\,mag cannot
be explained by ellipsoidal variations for such models. The fit of
the $B$ light curve corresponds to a larger $\chi_{\nu}^{2}$; the
corresponding $\chi_{\nu}^{2}$ would be similar to those found for
the data from the other filters provided that the photometric errors
used to normalize the $\chi_{\nu}^{2}$ were 0.010\,mag instead of
0.007\,mag.

\begin{table}
\caption{Fit of the light curves of LSS~3074 by a sine function. \label{tab_sinus}}

\begin{tabular}{r|cccc}
\hline 
 & \multicolumn{4}{c}{Filter}\tabularnewline
\hline 
 & $B$  & $V$  & $R$  & $I$\tabularnewline
\hline 
\hline 
$P=1.0926$ d &  &  &  & \tabularnewline
$a(P)$  & 0.0901  & 0.0910  & 0.0964  & 0.0980\tabularnewline
$\sigma{}_{a}$  & 0.0022  & 0.0022  & 0.0039  & 0.0027\tabularnewline
$\chi_{\nu}^{2}$  & 5.96  & 2.92  & 2.65  & 2.66\tabularnewline
$\sigma_{{\rm fit}}$  & 0.0171  & 0.0171  & 0.0212  & 0.0212\tabularnewline
\hline 
$P=2.1852$ d &  &  &  & \tabularnewline
$a_{1}(P)$  & 0.0118  & 0.0125  & 0.0107  & 0.0125\tabularnewline
$\sigma{}_{a_{1}}$  & 0.0020  & 0.0020  & 0.0038  & 0.0020\tabularnewline
$a_{2}(P/2)$  & 0.0896  & 0.0905  & 0.0958  & 0.0905\tabularnewline
$\sigma{}_{a_{2}}$  & 0.0019  & 0.0019  & 0.0037  & 0.0019\tabularnewline
$\chi_{\nu}^{2}$  & 4.60  & 2.20  & 2.38  & 1.30\tabularnewline
$\sigma_{{\rm fit}}$  & 0.0150  & 0.0148  & 0.0200  & 0.0148\tabularnewline
\hline 
$N$  & 120  & 122  & 58  & 118\tabularnewline
$\sigma$  & 0.007  & 0.010  & 0.013  & 0.013\tabularnewline
\hline 
\end{tabular}

\tablefoot{$P$ is the adopted period, $a{}_{i}$ the semi-amplitudes,
$\sigma{}_{a_{i}}$ the 1$\sigma$ error on semi-amplitude, $\chi_{\nu}^{2}$
the minimum reduced $\chi^{2}$, $\sigma_{{\rm fit}}$ the r.m.s of
the residuals, $N$ the number of points and $\sigma$ the expected
error.}
\end{table}

The $\chi_{\nu}^{2}$ of Table \ref{tab_sinus} are too large and
this anomaly could have several origins. One possibility is that the
above-mentioned photometric errors deduced from the comparison stars
might not apply to the case of LSS~3074. This could be due to an
unfortunate location on the CCDs or for example to strange behaviour
due to peculiar stellar colours. This interpretation is however not
very likely. Alternatively, the fitted mathematical model might not
be correct. Since any periodic function can be expanded into a Fourier
series and we found that our sine fits do not require additional terms,
the additional, non-periodic component of the model behaves similarly
to an observational noise although it must have a different origin.
A possible explanation could be an intrinsic variability of the star
at high frequencies since no low-frequency variations were previously
detected beyond the orbital one. Of course, this additional component
has an upper limit. Since the sine fits are our best models, we can
consider estimating this upper limit from the corresponding residuals.
In the $V$ filter for instance, we would need an additional variability
compared to the expected $\sigma=0.010$\,mag photometric error that
quadratically adds to it to reach $\sigma_{{\rm fit}}=0.0148$\,mag
for the two periods fit or $\sigma_{{\rm fit}}=0.0171$\,mag for
the single period fit. Considering that the difference in depth between
the two minima is not well established, we can adopt the latter and
the relevant value would thus be $\sigma_{{\rm fit}}=0.0171$\,mag.

The transition from $\sigma=0.010$ to $0.0171$\,mag translates
into a change from $\sigma=0.009$ to $\sigma^{{\rm corr}}=0.0185$\,mag
for the global fit of the data from all four filters. The substitution
by these newly estimated errors alleviates the rejection of the NIGHTFALL
model, since it reduces the $\chi{}^{2}$. However, this is purely
artificial. In addition, the $\chi_{\nu}{}^{2}$ of 4.4 could still
be due to a lack of ability of the NIGHTFALL model to fit the light
curve, even if it were noiseless. However, under the hypothesis of
the fit of the NIGHTFALL model, the errors on the parameters must
be modified to remain coherent. Even though this additional dispersion
does not stem from the same origin as the genuine photometric errors,
impact of this dispersion on the derived parameters is similar to
that of photometric errors.

To obtain the 1$\sigma$ error on the parameters of the NIGHTFALL
models, we have to consider the dispersion at $\chi_{{\rm min}}^{2}+\Delta\chi_{1\sigma}^{2}$
where $\Delta\chi_{1\sigma}^{2}=2.3$ for the 1$\sigma$ confidence
interval of the simultaneous adjustment of two free parameters ($i$
and $fill_{{\rm p}}$ = $fill_{{\rm s}}$ as is the case when we fit
the light curve assuming an overcontact configuration with NIGHTFALL).
From the values of the $\chi^{2}$ of our best-fit NIGHTFALL model,
the 1$\sigma$ error bars on the orbital inclination and the filling
factors would be unrealistically small (e.g.\ about $0.1^{\circ}$
for the error on $i$). However, as pointed out above, the dispersion
of the photometric data about the best-fit synthetic light curve exceeds
the value of our estimated photometric errors, suggesting that there
could be an intrinsic photometric variability in addition to the orbital
modulation.

Admitting the impact of the additional dispersion, we have to correct
the $\Delta\chi_{1\sigma}^{2}=2.3$ to take this effect into account.
Therefore, we have to adopt a corrected $\Delta\chi_{1\sigma,{\rm corr}}^{2}=\Delta\chi_{1\sigma}^{2}\times\left(\frac{\sigma^{{\rm corr}}}{\sigma}\right)^{2}=8.9$.
We have then estimated the errors on the NIGHTFALL model parameters
by accordingly adopting this new corrected value. In this way, rounding
the values uppards, we estimate errors of $1.0^{\circ}$ and of 0.01
for the inclination and the filling factors of both stars, respectively.

\section{Spectral analysis \textmd{\label{Modelatmosphere}}}

\subsection{Rotational velocities and macroturbulence}

In order to determine the projected rotational velocities ($v\,\sin{i}$)
of the stars of the system, we applied a Fourier transform method
(Sim\'on-D\'iaz \& Herrero \cite{Simon-Diaz}, Gray \cite{Gray}). For
the primary star, we used the profiles of the He\,\textsc{i} $\lambda$\,4471,
He\,\textsc{ii} $\lambda\lambda$\, 4200, 4542, 6118, and N\,\textsc{iii}
$\lambda$\,6075 lines of the reconstructed spectrum. We selected
these lines as they are rather well isolated in the spectra and should
thus be free of blends. Unfortunately, we could not use the same lines
for the secondary star because they are particularly deformed. Indeed,
the red wing of these lines is steeper than the blue wing. We then
used the He\,\textsc{ii} $\lambda$\,5412 and Si\,\textsc{iv} $\lambda$\,4089
lines to determine the projected rotational velocity of the secondary
star. The results are presented in Table\,\ref{vsini}. The mean
$v\,\sin{i}$ of the primary star is $(110\pm13)$\,km\,s$^{-1}$,
while that of the secondary is $(127\pm6)$\,km\,s$^{-1}$. Owing
to the small number of lines usable to study the rotation of the secondary
star, we could not use a classical standard deviation method to determine
the associated error bars. We thus studied the Fourier transforms
for a sample of rotational velocities to estimate these errors. From
these values of $v\,\sin{i}$ and considering the inclination of the
system from Sect.\,\ref{photometry}, we found that the stars are
in synchronous rotation with a rotational period very close to the
orbital period of the system.

\begin{table}[h]
\caption{Projected rotational velocities ($v\,\sin{i}$ expressed in km\,s$^{-1}$)
of the components of LSS~3074.}

\label{vsini}

\centering{}%
\begin{tabular}{c|cc}
\hline 
Line  & Primary  & Secondary\tabularnewline
\hline 
\hline 
Si\,\textsc{iv} $\lambda$\,4089 & - & 126\tabularnewline
He\,\textsc{ii} $\lambda$\, 4200 & 115 & -\tabularnewline
He\,\textsc{i} $\lambda$\,4471 & 97 & -\tabularnewline
He\,\textsc{ii} $\lambda$\, 4542 & 119 & -\tabularnewline
He\,\textsc{ii} $\lambda$\,5412 & - & 128\tabularnewline
N\,\textsc{iii} $\lambda$\,6075 & 92 & -\tabularnewline
He\,\textsc{ii} $\lambda$\,6118 & 125 & -\tabularnewline
\hline 
Mean value  & 110 $\pm$ 13  & 127 $\pm$ 6\tabularnewline
\hline 
\end{tabular}
\end{table}

Macroturbulence is defined as a non-thermal motion in the stellar
atmosphere with turbulent cells larger than the mean free-path of
the photons. The main effect of macroturbulence is an additional broadening
of the spectral lines. An approximation of the macroturbulence velocities
was obtained by applying the radial-tangential anisotropic macroturbulent
broadening formulation of Gray (\cite{Gray}) on the spectra, after
the inclusion of rotational velocity broadening. For this purpose,
we used the auxiliary program MACTURB of the stellar spectral synthesis
program SPECTRUM v2.76 developed by Gray (\cite{macturb}). We applied
this technique to the lines \ion{He}{i} $\lambda\lambda$ 4026,
4471, and 5016, and \ion{He}{ii} $\lambda$ 4542 and obtained
macroturbulence velocities of 20 and 50\,km\,s$^{-1}$, for the
primary and secondary stars, respectively.

\medskip{}

Both rotational and macroturbulence velocities were applied on the
synthetic spectra (see Sect.\,\ref{CMFGENcode}) before comparing
the latter with the disentangled spectra.

\subsection{CMFGEN code \label{CMFGENcode}}

In order to determine the fundamental properties of both components
of LSS\ 3074, we used the non-LTE model atmosphere code CMFGEN (Hillier
\& Miller \cite{HM}). This code solves the equations of radiative
transfer and statistical equilibrium in the co-moving frame, is designed
to work for both plane-parallel and spherical geometries, and can
be used to model Wolf-Rayet stars, O stars, luminous blue variables
and supernovae. A super-level approach is adopted for the resolution
of the equations of statistical equilibrium. The CMFGEN code also
further accounts for line blanketing and its impact on the energy
distribution. The hydrodynamical structure of the stellar atmosphere
is directly specified as an input to the code, and a $\beta$ law
is used to describe the velocity law within the stellar winds. We
included the following chemical elements and their ions in our calculations
: H, He, C, N, O, Ne, Mg, Al, Si, S, Ca, Fe, and Ni. The solution
of the equations of statistical equilibrium is used to compute a new
photospheric structure, which is then connected to the same $\beta$
wind velocity law. The radiative transfer equations were solved based
on the structure of the atmosphere with a microturbulent velocity
varying linearly with wind velocity from 20 km s$^{-1}$ in the photosphere
to $0.1$ \texttimes{} v$_{\infty}$ at the outer boundary, and generated
synthetic spectra were compared to the reconstructed spectra of the
primary and secondary stars.

\medskip{}

As a first approximation, the mass-loss rates and $\beta$ parameters
were taken from Muijres et al.\ (\cite{Muijres}) for the spectral
types of both stars, whilst the wind terminal velocity was assumed
equal to the mean value for stars of the same spectral type (Prinja
et al.\ \cite{Prinja}). We then estimated the surface gravities,
stellar masses, radii, and luminosities from our study of the photometric
curves. The relevant parameters were then adjusted via an iterative
process, as each adjustment of a given parameter leads to some modifications
in the value of other parameters. This process and the results we
obtained are presented in subsections \ref{Method} and \ref{Results},
respectively. As explained in Sect. \ref{lightcurve}, this iterative
process was then coupled with a second iterative process, through
the study of the photometric light curve.

\subsection{Method\label{Method}}

The first step is to adjust the effective temperature in the models
of the stars. To do so, we adjusted the relative strengths of the
He\,\textsc{i} $\lambda$\,4471 and He\,\textsc{ii} $\lambda$\,4542
lines (Martins \cite{Martins}). Final values are 39 900 and 34 100
K for the primary and secondary stars, respectively. Once the effective
temperatures were determined, we used them, together with the stellar
radii obtained in our study of the photometric curves, to constrain
the luminosities of the stars. We therefore obtained luminosities
of $1.38\times10^{5}L_{\odot}$ and $8.49\times10^{4}L_{\odot}$ for
the primary and secondary stars.

Subsequently, the binarity of the studied system causes some problems.
Indeed, the next step is to adjust the surface gravities in the models,
but as we mentioned in Sect.\,\ref{disent}, the wings of the Balmer
lines are too broad to be properly recovered by the spectral disentangling.
To circumvent this problem, we recombined the models of the primary
and secondary star spectra for several phases and compared the resulting
binary spectra directly to the observations. This permitted us to
better adjust the gravities, but with larger uncertainties than in
the case of a single star.

A second problem arises with the adjustment of the wind parameters.
Indeed, the He\,\textsc{ii} $\lambda$\,4686 and Balmer lines may
be polluted by some emission from the wind-wind interaction zone.
Therefore, the uncertainties on the terminal velocity, the $\beta$
of the velocity law, the clumping factor, the clumping velocity factor,
and the mass-loss rate are quite high. Indeed, considering the possible
existence of such additional emission polluting these diagnostic lines,
the adjustment of the models onto the reconstructed spectra may lead
to an underestimate of the terminal velocity, and an overestimate
of the $\beta$ of the velocity law, clumping factor, clumping velocity
factor, and mass-loss rate. Therefore, the obtained values can only
be considered as lower and upper limits of the real properties of
the stellar winds. However, we still adjusted them as well as possible:
we obtained for the primary and secondary stars values of 2615 and
3055 kms$^{-1}$ for $v{}_{\infty}$, 1.30 and 1.40 for $\beta$ and
$3.00\times10^{-6}$ and $3.51\times10^{-6}$ M$_{\odot}$\,yr$^{-1}$
for the mass-loss rate, based on the strength of H$\alpha$, the width
of the He\,\textsc{ii} $\lambda$\,4686 and H$\alpha$ lines, and
both the strengths of H$\gamma$ and H$\delta$, respectively. The
clumping formalism used in the CMFGEN model is

\begin{center}
$f(r)=f_{1}+(1\lyxmathsym{\textminus}f_{1})e^{(\lyxmathsym{\textminus}\frac{V(r)}{f_{2}})}$ 
\par\end{center}

\noindent with a clumping filling factor $f_{1}$ of 0.7, a clumping
velocity factor $f_{2}$ of 200 kms$^{-1}$ and $V(r)$ the velocity
of the wind for both primary and secondary stars, and we adjusted
these two clumping factors through the strength and shape of H$\alpha$
and H$\beta$ lines.

Finally, once the fundamental properties of the stars had been established
or fixed, we investigated the CNO abundances within their atmosphere
through the strengths of associated lines. At that point, we encountered
several problems. Indeed, there is no visible O line in both primary
and secondary spectra, and the only present C lines are the \ion{C}{iv}
$\lambda\lambda$\,5801 and 5812 lines, which cannot be considered
in the determination of the surface C abundance since they can exhibit
complex profiles influenced by various phenomena among which is continuum
fluorescence (Conti \cite{Conti2}; Bouret et al.\ \cite{Bouret}).
We thus were only able to determine upper limits for C and O abundances.
We used the \ion{N}{iii} $\lambda$\,4379 and \ion{N}{iii}
$\lambda\lambda$\,4511-15-18-24-30-35 lines to adjust the N abundance
for the primary star. The same lines were used for the secondary star
except for the \ion{N}{iii} $\lambda$\,4530 and \ion{N}{iii}
$\lambda$\,4535 lines. These N abundance diagnostic lines are taken
from Martins et al.\ (\cite{Martins3}) study of a sample of 74 objects
comprising all luminosity classes and spectral types from O4 to O9.7.
We selected, from their study, all the suggested lines that were significantly
seen in our observations. We performed a normalized $\chi^{2}$ analysis
to determine the best fit to these lines (Martins et al.\ \cite{Martins3}).
The normalization consists in dividing the $\chi^{2}$ by its value
at minimum, $\chi\text{\texttwosuperior}{}_{{\rm }{min}}$. As a 1$\sigma$
uncertainty on the abundances, we then considered abundances up to
a $\chi^{2}$ of 2.0, i.e.\ an approximation for 1 over $\chi\text{\texttwosuperior}{}_{{\rm {min}}}$,
as suggested by Martins et al.\ (\cite{Martins3}).

\subsection{Results\label{Results}}

Figure\,\ref{fig1} shows the best fit of the optical spectra of
the primary and secondary stars obtained with CMFGEN, and we present
in Table\,\ref{CMFGENparam} the associated stellar parameters.

Figure\,\ref{fig1} reveals that we encountered a lot of difficulty
adjusting the spectra of the components of LSS 3074. First of all,
it clearly shows that the noise is strongly enhanced in the reconstructed
secondary spectrum, because of the high brightness ratio of the system.
This peculiarity makes the adjustment of the secondary spectrum even
more difficult. Second, if we can see that the He lines are generally
well reproduced for both stars, some of these lines in the observed
spectra, especially for the secondary star, feature some emission
in their wings that is not reproduced by the models. We can also see
that the Balmer and He\,\textsc{ii} $\lambda$\,4686 lines are not
perfectly well recovered, probably owing to the complexity of the
winds and the presence of a wind-wind interaction zone, as discussed
in Sect.\,\ref{Method}.

\begin{figure}[htb]
\begin{centering}
\textbf{\resizebox{8.5cm}{!}{\includegraphics{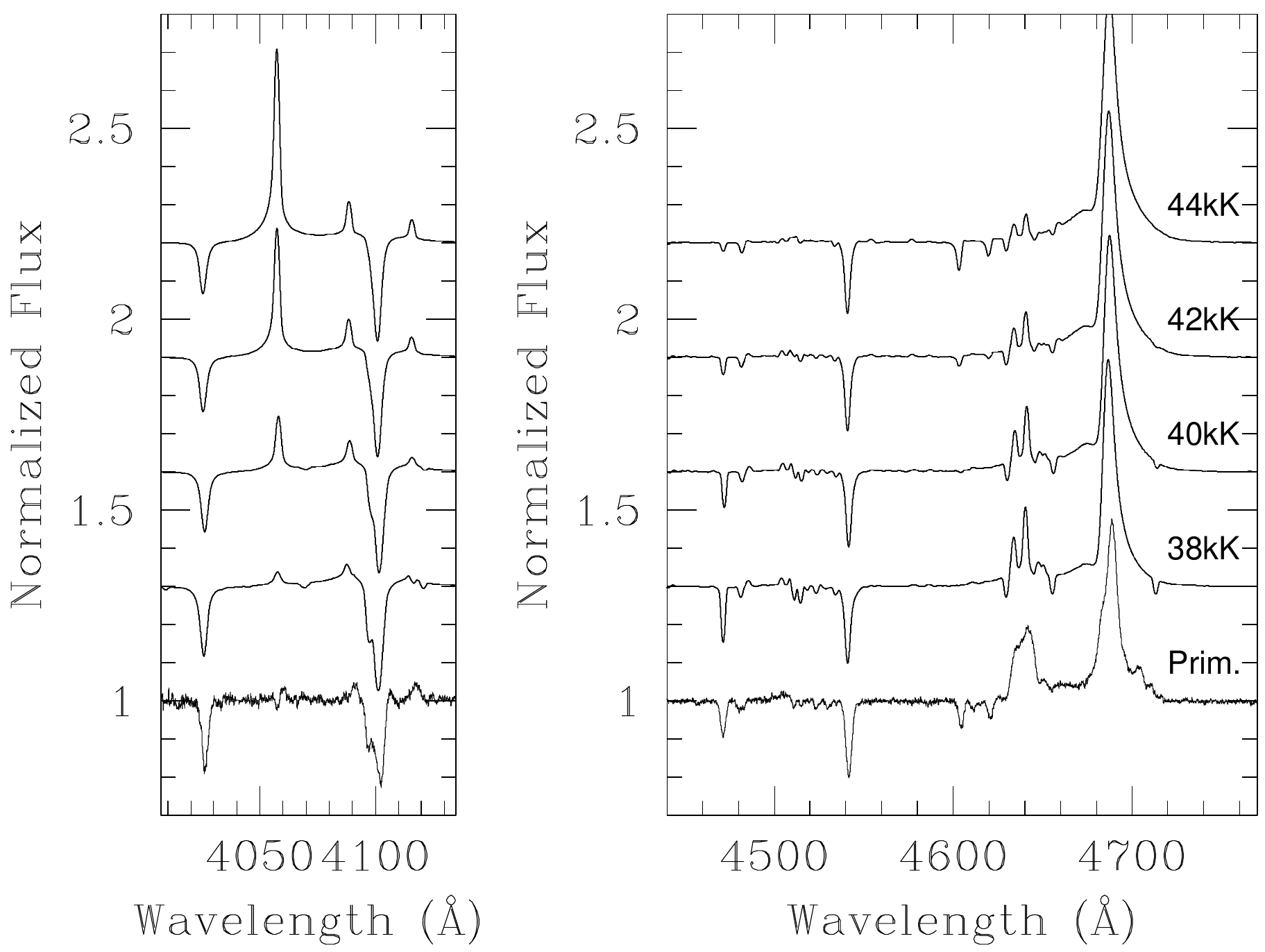}} }
\par\end{centering}

\textbf{
\caption{Evolution of the strength of the \ion{N}{iii} $\lambda\lambda$\,4634-41,
\ion{N}{iv} $\lambda$\,4058, and \ion{N}{v} $\lambda\lambda$\,4604,
4620 lines in synthetic CMFGEN spectra as a function of effective
temperature. The primary spectrum of LSS~3074 obtained via disentangling
is shown at the bottom for comparison. \label{figN}}
}
\end{figure}

\begin{figure*}[!t]
\includegraphics[width=8.8cm]{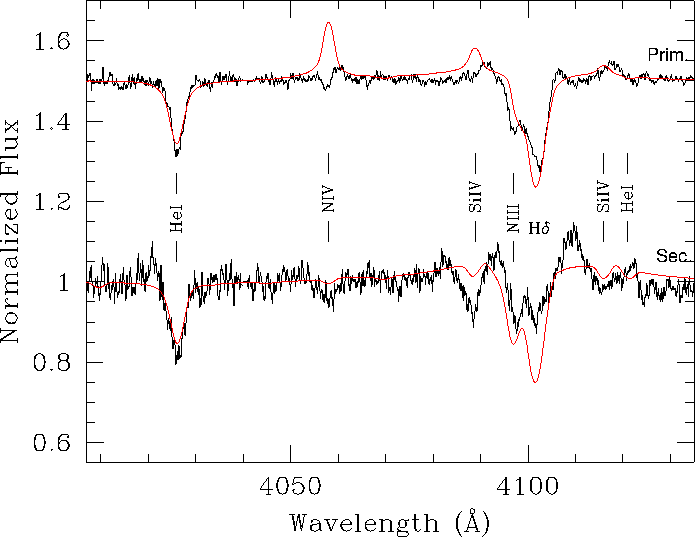}\hspace{0.4cm}\includegraphics[width=8.8cm]{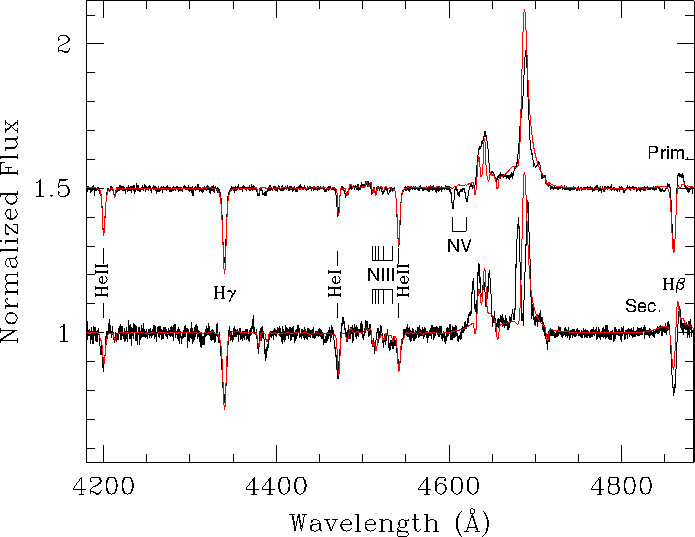}

\includegraphics[width=8.75cm]{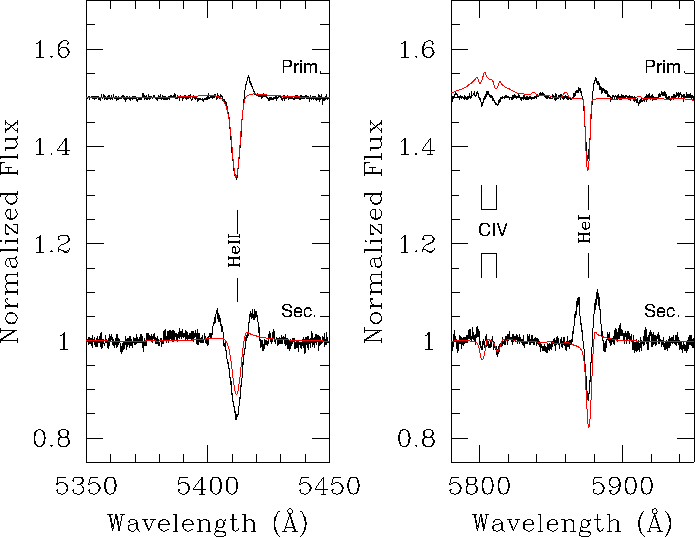}\includegraphics[width=9.25cm]{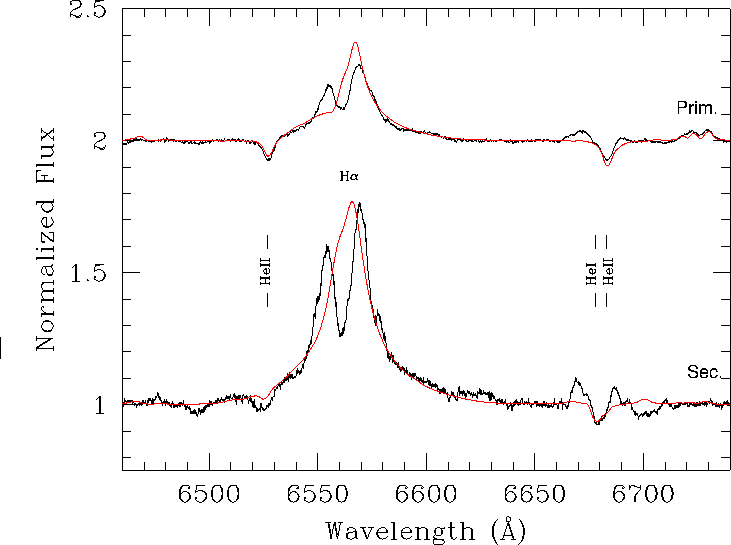}
\caption{Part of the normalized separated spectra of the primary (top, shifted
upwards by 0.5 continuum units in the first three panels, and by 1.0
continuum units for the last panel) and secondary stars (bottom),
along with the best-fit CMFGEN model spectra (red). \label{fig1}}
\end{figure*}

Concerning the N lines, most \ion{N}{iii} features are reasonably
well reproduced by the models, but a few lines of other nitrogen ions
are not, especially in the primary spectrum. Indeed, the model predicts
a rather strong \ion{N}{iv} $\lambda$\,4058 emission, while
the observations clearly show that this is not the case. Another remarkable
issue is the presence in the observed primary spectra of the \ion{N}{v}
$\lambda\lambda$\,4604, 4620 lines, which are not predicted by the
model. These \ion{N}{v} lines are seen in the spectra of the
earliest O supergiants (starting from O4\,{}If Walborn et al.\ \cite{Walborn02},
Sota et al.\ \cite{Sota11}) and, in some cases, in the spectra of
early O main-sequence stars (e.g.\ Mahy et al.\ \cite{Mahy}), and
could suggest that our value of the primary effective temperature
(39\,900\,K) is too low. To test this hypothesis, we computed a
set of CMFGEN models with temperatures ranging from 38\,000 to 44\,000\,K.
As one can see on Fig.\,\ref{figN}, a temperature of about 44\,000\,K
would be needed to reproduce the strength of the \ion{N}{v} lines.
However, in this case, the model predicts \ion{N}{iii} emissions
that are much weaker than observed as well as a huge \ion{N}{iv}
line that is clearly not observed. To reproduce the strength of the
observed \ion{N}{iii} and \ion{N}{iv} lines, a lower temperature
of 38\,000\,K seems more indicative. But this lower temperature
is clearly not sufficient to produce the \ion{N}{v} lines. The
temperature that we determined in our CMFGEN fits is the temperature
that properly reproduces the strength of the helium lines and represents
a compromise between the contradicting trends of the nitrogen lines. 

\textbf{
}

Our difficulty in reproducing simultaneously the lines of \ion{N}{iii},
\ion{N}{iv}, and \ion{N}{v} in the spectra of O supergiants
is actually not new. Crowther et al.\ (\cite{Crowther}) and Rauw
et al.\ (\cite{HDE228766}) encountered similar problems in their
attempts to fit the spectra of the O4\,Iaf$^{+}$ star HDE~269\,698
and of the O7 + O4\,If/WN8ha binary HDE~228\,766, respectively.
More recently, Bouret et al.\ (\cite{Bouret}) applied the CMFGEN
code to a sample of early O supergiants. These authors also failed
to reproduce consistently the lines of the three nitrogen ions in
the spectra of the three O4\,If supergiants HD~15\,570, HD~16\,691,
and HD~190\,429A for which they derived effective temperatures of
38\,000, 41\,000, and 39\,000\,K, respectively. We note that these
effective temperatures are very close to our best-fit value for the
primary of LSS~3074.

The origin of these difficulties is currently unclear. One possibility
could be that the models do not account (correctly) for the Auger
effect. X-ray emission arising in the wind and back illuminating the
stellar photosphere could lead to an overionization of nitrogen and
favour the formation of the N \textsc{v} lines. In the specific case
of the primary of LSS~3074, the simultaneous presence of the three
ionization stages of nitrogen could also be from a non-uniform surface
temperature as a result of gravity darkening. Applying the von Zeipel
theorem to the primary, we find that the effective temperature near
the poles should be about 15\% higher than near the equator. The observed
spectrum of the primary results from the combination of the spectra
emitted by the various regions, which have their own temperatures
(Palate \& Rauw \cite{Palate12}, Palate et al.\ \cite{Palate13}).
Therefore, gravity darkening could indeed result in a spectrum with
an unusual combination of spectral lines.

\begin{table}
\caption{Best-fit CMFGEN model parameters of the primary and secondary stars.\label{CMFGENparam}}

\begin{centering}
\begin{tabular}{c|cc}
\hline 
\multicolumn{1}{c|}{} & Prim.  & Sec. \tabularnewline
\hline 
\hline 
$R$ (R$_{\odot}$)  & $7.5\pm0.6$ & $8.2\pm0.7$\tabularnewline
$M$ (M$_{\odot}$)  & $14.6\pm2.1$ & $17.2\pm3.0$\tabularnewline
$T_{{\rm eff}}$ ($10^{4}$\,K)  & $3.99\pm0.15$ & $3.41\pm0.15$\tabularnewline
log(\emph{g}) (cgs) & $3.82\pm0.20$ & $3.83\pm0.20$\tabularnewline
log ($\frac{L}{L_{\odot}}$) & $5.14\pm0.07$ & $4.93\pm0.08$\tabularnewline
$\beta$  & $\leq$1.30 & $\leq$1.40\tabularnewline
$v_{\infty}$ (km\,s$^{-1}$)  & $\geq$2615 & $\geq$3055\tabularnewline
$\dot{M}/\sqrt{f_{1}}$ (M$_{\odot}$\,yr$^{-1}$)  & $\leq$$3.00\times10^{-6}$ & $\leq$$3.51\times10^{-6}$\tabularnewline
\hline 
\end{tabular}
\par\end{centering}

\tablefoot{The quoted errors correspond to 1$\sigma$ uncertainties.} 
\end{table}

\begin{table}
\caption{Chemical abundances of the components of LSS~3074.}

\resizebox{9cm}{!}{%
\begin{tabular}{c|ccc}
\hline 
\multicolumn{1}{c|}{} & Primary  & Secondary  & Sun\tabularnewline
\hline 
\hline 
He/H  & 0.25 & 0.09 & 0.089\tabularnewline
C/H  & $\leq6.45\times10^{-5}$ & $\leq2.05\times10^{-5}$ & $2.69\times10^{-4}$\tabularnewline
N/H  & $5.65{}_{-3.02}^{+4.07}\times10^{-4}$  & $3.69_{-0.91}^{+0.83}\times10^{-4}$  & $6.76\ensuremath{\times10^{-5}}$\tabularnewline
O/H  & $\leq2.67\times10^{-5}$ & $\leq1.00\times10^{-5}$ & $4.90\times10^{-4}$\tabularnewline
\hline 
\end{tabular}}

\tablefoot{Abundances are given by number as obtained with CMFGEN.
The solar abundances (\cite{Asplund}) are quoted in the last column.
The 1$\sigma$ uncertainty was set to abundances corresponding to
a normalized $\chi^{2}$ of 2.0.} \label{table4} 
\end{table}

However, we can stress three very interesting results from our spectral
analysis. First of all, as we can see in Table\,\ref{CMFGENparam},
we confirm the very low masses and radii predicted by Morrell \& Niemela
(\cite{Morrell}), which is unusual for stars with such spectral types.
We can also see that the system seems to display strong winds, even
though these winds are not perfectly constrained. Finally, we found
a strong overabundance in N and a strong depletion in C and O in both
stars of the system, together with a strong He enrichment of the primary
atmosphere.

\section{Discussion and conclusion\textmd{s\label{Conclusions}}}

Our analysis of the light curve of LSS~3074 indicates that this system
is a candidate for an overcontact binary. Massive overcontact binaries
are important objects as they could be progenitors of merger events
or of binary black hole systems (Lorenzo et al.\ \cite{MYCam}, Almeida
et al.\ \cite{Almeida}). Yet, only a handful of such systems are
known and sometimes the exact configuration (double contact versus
semi-detached) is controversial (e.g.\ Ya\c{s}arsoy \& Yakut \cite{V382Cyg}
and Zhao et al.\ \cite{LYAur}, respectively). Among the best-studied
systems hosting O-type stars are HD~100\,213 (= TU Mus, Linder et
al.\ \cite{Linder07}), VFTS~352 (Almeida et al.\ \cite{Almeida}),
and MY~Cam (= BD+56\textdegree{}864, Lorenzo et al.\ \cite{MYCam}).
Lorenzo et al.\ (\cite{MYCam}) and Almeida et al.\ (\cite{Almeida})
note that the components of MY~Cam and VFTS~352 appear hotter than
one would expect from their masses. This discrepancy also exists for
LSS 3074 and is even worse than in the other systems. Lorenzo et al.\ (\cite{MYCam})
interpret this result as a consequence of the highly elongated shape
of the stars. In the case of VFTS~352, Almeida et al.\ (\cite{Almeida})
further report an enhanced He abundance. These authors interpret this
situation as the result of enhanced mixing in a tidally locked binary
(de Mink et al.\ \cite{deMink}) and suggest that the system is currently
in a long-lived overcontact phase of case A mass transfer. Again LSS
3074 also shows abundance anomalies indicating that both stars are
evolved (enhanced N abundance) and the primary is the more evolved
star (enhanced He abundance).

\subsection{Evolutionary status\label{Evolutionary-status}}

We first remark that we did not consider the possibility of quasi-chemically
homogeneous evolution as described in Heger et al.\ (\cite{Heger}),
because the observed rotational velocities of both components are
too small. 

The altered CNO abundances in the atmosphere of the primary, the luminosity
of primary, and the increased helium abundance are strong arguments
that the star is in the slow phase of case B RLOF. If the star were
post-RLOF, accounting for the present mass, its luminosity would have
to be log ($\frac{L}{L_{\odot}}$) = 5.5 rather than the observed
5.14 (e.g.\ Vanbeveren et al.\ \cite{VDLVR}, formula 5.4). Moreover,
the observed CNO abundances of the secondary seem to indicate that
at least some mass lost by the primary was accreted by the secondary.
The normal (solar) helium abundance, on the other hand, lead us to
conclude that not all the helium enhanced layers lost by the primary
were accreted by the secondary. Accounting for these facts a theoretical
study can be undertaken to attempt to identify its possible progenitors.
For this purpose, the Brussels binary population synthesis code is
used. This code uses as input thousands of detailed evolutionary calculations
performed with the Brussels binary evolution code. Both codes are
described in detail in De Donder \& Vanbeveren (\cite{DeDonder}).
However, the theoretical treatment of binary evolution is still subject
to a number of uncertainties, which are typically implemented by means
of parameters. For the present discussion, two uncertain aspects of
binary mass transfer, characterized in population synthesis by two
parameters, are of interest. The first is the fraction of the mass
lost by the donor star that is actually accreted by the gainer star.
This fraction necessarily lies between zero (i.e.\ totally non-conservative
mass transfer) and one (i.e.\ totally conservative mass transfer),
and is characterized in population synthesis by the parameter $\beta_{{\rm RLOF}}$%
\footnote{The usual designation for this parameter is $\beta$, but we used
the $\beta_{{\rm RLOF}}$ designation to avoid any confusion with
the $\beta$ of the velocity law in the stellar winds.%
} (hence $0\leq\beta_{{\rm RLOF}}\leq1$). If $\beta_{{\rm RLOF}}<1$,
and thus mass is lost from the system, a second critical assumption
concerns how much angular momentum is taken along from the system
by this lost mass. This is determined by a parameter $\eta$ (for
a formal definition, see De Donder \& Vanbeveren \cite{DeDonder}).
For the present discussion, it suffices to know that the larger $\eta$,
the more angular momentum is removed from the binary by a given amount
of mass loss. A common assumption made in many population synthesis
studies is that the mass lost takes along only the specific orbital
angular momentum of the gainer star (and thus leaves this star in
a spherically symmetric way). This results in a low angular momentum
loss, characterized by a low value of $\eta \sim 0.05$. An alternative
assumption is that mass is lost through the second Lagrangian point
and forms a circumbinary ring, taking along some of the orbital angular
momentum of this ring. This entails a much larger angular momentum
loss, and has been shown to result in $\eta=2.3$. Various other,
less common assumptions are also possible (e.g.\ gainer orbital plus
spin angular momentum and specific binary angular momentum), which
all result in intermediate cases and can be approximated by taking
$\eta=1$. The purpose of this paragraph is to investigate, for various
combinations of the parameters $\beta_{{\rm RLOF}}$ and $\eta$,
whether they can indeed result in the formation of a binary in the
slow phase of case B RLOF with the physical parameters (masses and
orbital period) of LSS3074, and if so, originating from which initial
conditions. All calculations are performed for Solar-like metallicity.

\medskip{}

Table\,\ref{tab_Models} shows the initial masses and orbital period,
depending on the assumed values of $\beta_{{\rm RLOF}}$ and $\eta$,
which produce a 14.8+17.2 M$_{\odot}$ binary in the slow phase of
case B RLOF and with an orbital period of 2.2 days.

\begin{table*}
\centering
\caption{Initial masses and orbital period, leading to a 14.8+17.2 M$_{\odot}$, 2.2 d post-RLOF binary for different combinations of $\beta_{\rm{RLOF}}$ and $\eta$.}\label{tab_Models}
\begin{tabular}{c | c c c}
\hline 
&$\beta_{\rm{RLOF}}=1$ & $\beta_{\rm{RLOF}}=0.5$ & $\beta_{\rm{RLOF}}=0.1$ \\
\hline
$\eta=2.3$  &   no solution	&  34+7.9 M$_{\odot}$, 560 d	&  34+15 M$_{\odot}$, 1000 d	\\ $\eta=1$	&   no solution	&  34+7.9 M$_{\odot}$, 90 d	&  34+15 M$_{\odot}$, 80 d	\\ 
$\eta \sim 0.05$    &	no solution  &  34+7.9 M$_{\odot}$, 5.3 d	&  34+15 M$_{\odot}$, 1.4 d	\\
\hline
\end{tabular}
\end{table*}

Our simulations reveal that the primary progenitor was a star with
initial mass 30-35 M$_{\odot}$. We find no physically realistic solution
if the RLOF is treated conservatively. If it is assumed that $\beta_{\rm{RLOF}}=0.5$,
we have to start with an initial 34+7.9 M$_{\odot}$ binary. It is
however doubtful that a binary with such an extreme initial mass ratio
would avoid merging. Our best guess is then a model with $\beta_{\rm{RLOF}}<0.5$,
such as the $\beta_{\rm{RLOF}}=0.1$ model of Table\,\ref{tab_Models}.
If the matter that leaves the binary takes with it the specific angular
momentum of the gainer ($\eta=0.05$), the initial period of the binary
had to be very small (of the order of 1 or 2 days) and this means
that the progenitor binary first went through a case A RLOF phase
followed by case B. Interestingly, our simulations show that the slow
case B RLOF will stop after some 10000 yrs. The primary star then
will be a 11-12 M$_{\odot}$ WR star, i.e.\ the binary will be very
similar to the WR binary CQ Cep.

\subsection{Summary and conclusions}

From what we have seen above, LSS~3074 is clearly a very peculiar
binary system that challenges our current methods of binary spectral
analysis. In this paper, we have tried to get a deeper understanding
of this system, but we encountered a number of problems that prevent
us from establishing a fully consistent explanation of all the properties
of the system. Here we discuss these problems and try to highlight
some avenues for possible solutions.

The properties of the components of LSS~3074 that we determined from
our spectral and photometric analysis do not concur with those of
genuine early O-type supergiants as determined, for instance, by Bouret
et al.\ (\cite{Bouret}). Indeed, these latter authors used the CMFGEN
code to analyse UV and optical spectra of typical early and mid-O
supergiants. They found that O4-4.5 supergiants have radii between
18.5 and 21.6\,R$_{\odot}$, $\log{g}$ in the range 3.51 to 3.66
and $\log{L/L_{\odot}}$ between 5.83 and 5.96. For O6-7.5 supergiants,
Bouret et al.\ (\cite{Bouret}) determined radii between 20.5 and
21.3\,R$_{\odot}$, $\log{g}$ between 3.41 and 3.54, and $\log{L/L_{\odot}}$
in the range 5.68 to 5.80. The components of LSS~3074 have significantly
smaller radii, which are actually even smaller than the typical radii
of early-O main-sequence stars. Stars as big as the supergiants analysed
by Bouret et al.\ (\cite{Bouret}) would not fit into such a short-period
binary system as LSS~3074, even considering, as we have here, the
possibility of an overcontact configuration. The luminosities of the
components of LSS~3074 are also far from those determined by Bouret
et al.\ (\cite{Bouret}), and are again lower than those of main-sequence
stars of the same spectral type. As for the $\log{g}$ of the stars
in LSS~3074, their values are higher than those of genuine supergiants,
indicating a more compact stellar atmosphere than for supergiants.
The values we found are intermediate between those of giants and main-sequence
stars. Last but not least, the masses of the components (14.6 and
17.2\,M$_{\odot}$ for the primary and secondary, respectively) are
considerably lower than the spectroscopic or evolutionary masses of
genuine O supergiants, which are around 50\,M$_{\odot}$ for early-O
supergiants and around 40\,M$_{\odot}$ for mid-O supergiants (Bouret
et al.\ \cite{Bouret}). In fact, the masses that we determined are
more compatible with those of O9-B0 main-sequence stars.

Yet, despite these very discrepant stellar properties, the spectrum
of the primary star of LSS~3074 is very similar to the spectrum of
HD~16\,691 (O4\,If) analysed by Bouret et al.\ (\cite{Bouret}),
including the presence of N\,\textsc{v} $\lambda\lambda$\,4605-4620
in absorption and the weakness of N\,\textsc{iv} $\lambda$\,4058.
This situation suggests that the spectra of the LSS~3074 could be
biased in some way. Our spectral analysis relies on the individual
spectra reconstructed via spectral disentangling. Such a reconstructed
spectrum is a brightness-weighted mean over the various parts of the
stellar surface that can be observed given the inclination of the
system. Owing to the effect of gravity darkening, the temperature
at the surface of a rotating star is not uniform: the polar regions
that have the higher value of $\log{g}$ are hotter than the equatorial
regions where $\log{g}$ is lower. Things are even more complicated
in a close binary system, where the local acceleration of gravity
is determined by the gradient of the Roche potential $\nabla\,\Omega$
(e.g.\ Palate \& Rauw \cite{Palate12}, Palate et al.\ \cite{Palate13}).
To evaluate whether or not this situation leads to a surface gravity
that comes close to that of a supergiant, we computed $|\nabla\,\Omega|$
over the surface of the primary star, adopting a Roche lobe filling
factor of 1.008. We then compared this value to the acceleration of
gravity one would have for a single star of same mass and radius.
Except for a small region near the bridge of material that connects
both stars, $|\nabla\,\Omega|$ is typically between 3 and 9 times
larger than the $g$ value for a single star. Therefore, the Roche
potential should lead to an increase of $\log{g}$ compared to its
value if binarity is neglected. Hence, this situation cannot explain
the discrepancy between the stellar properties and the observed spectrum.
A possible way out could be the effect of the radiation pressure on
the potential. Howarth (\cite{Howarth}) has shown that radiation
pressure leads to a scaling of the potential and a reduction of the
local gravity by a factor $1-\Gamma$, where $\Gamma$ is a measure
of the relative importance of radiation pressure over gravity. The
primary star of LSS~3074 could have lost a substantial fraction of
its initial mass to the surroundings and, as a result, appears now
as a very hot object, significantly hotter than expected from its
mass. This assumption is strengthened by the He and CNO primary surface
abundances, as shown in Sect.\,\ref{Evolutionary-status}. Given
the high temperature of the components of LSS~3074, the resulting
strong radiation pressure appears then to be the best candidate to
explain why the stars display unusual spectra.

Another odd feature of LSS~3074 is the discrepancy between the brightness
ratios inferred from spectroscopy and photometry. The spectroscopic
brightness ratio mainly stems from the weakness of the secondary spectral
lines. One possibility could be that these lines are filled in by
emission either from the primary star or from circumstellar material.
Some evidence for circumstellar emission can be seen in the He\,\textsc{i
$\lambda$\,5876 }and He\,\noun{ii} $\lambda$\,5412 lines of the
recontructed spectra, which are flanked by emission wings in the case
of the secondary. Yet, it seems unclear whether this effect applies
to all spectral lines and would be sufficient to explain the discrepancy
between photometry and spectroscopy.

The fact that the light curve of LSS~3074 is actually better fitted
by a simple sine function than by the NIGHTFALL model suggests that
the components of the binary system might actually be more elongated
than expected from the Roche potential. The inclusion of radiation
pressure in the calculation of the surface shape can actually lead
to systems that are detached whilst they would be in a contact configuration
based only on the Roche potential (Drechsel et al.\ \cite{Drechsel95},
Palate et al.\ \cite{Palate13}). Radiation pressure also changes
the ellipticity of the stars, making them appear less elongated for
a given value of the polar radius (e.g.\ Fig.~3 Palate et al.\ \cite{Palate13}),
although a further increase in the polar radius could lead to a higher
ellipticity whilst preserving a detached configuration (Drechsel et
al.\ \cite{Drechsel95}). Whilst a detailed modelling of the light
curve of LSS~3074, including the effect of radiation pressure, is
beyond the scope of the present study, we note that the temperature
distribution at the surface of the stars will also be altered and
this could possibly help explain the observed light curve. A detached
configuration resulting from the effect of radiation pressure could
in principle also help bring the photometric and spectroscopic brightness
ratios into better agreement.

\medskip{}

\begin{acknowledgements}
The Li\`ege team acknowledges support from the Fonds de la Recherche
Scientifique (F.R.S/FNRS) including especially an F.R.S/FNRS Research
Project (T.0100.15), as well as through an ARC grant for Concerted
Research Actions financed by the French Community of Belgium (Wallonia-Brussels
Federation), and an XMM PRODEX contract (Belspo). We are particularly
grateful to Reinhold Haefner for having made his IBVS data available
and to Ella Antokhina and Igor Antokhin for discussions on the photometric
solution. EG is greatly indebted to Stephanie Wachter and Suzanne
Tourtelotte for having spared no pains in facilitating the interaction
with the YALO archives. We are grateful to I.~Stevens, H.~Sana,
Y.~Naz\'e, J.-M.~Vreux, and M.~da Silva Pires for taking some of
the spectroscopic data or for discussions on earlier versions of this
work. We thank John Hillier for making his code available.\end{acknowledgements}

\noindent \medskip{}

\noindent \begin{flushleft}
\appendix{\bf{Appendix A}}{}
\par\end{flushleft}

\medskip{}

During the above-mentioned iterative process to derive the equivalent
of global differential magnitudes (see Sect.\,\ref{Photometry}),
a few stars were rejected owing to suspected variability. Two stars
turned out to be significantly variable: they are analysed here below. 

\medskip{}

\medskip{}

\noindent \begin{flushleft}
\appendix{\it{A.1. }}{\it{HD 116827 $\equiv$ CPD -61$\degr$3706}}
\par\end{flushleft}

\medskip{}

\noindent HD\,116827 is known to have H$\alpha$ in emission and
is thus classified as a Be star (MacConnell \cite{maco81}). The H$\beta$
line also presents a weak emission within the absorption profile;
the spectral type is B2Ve-B5Ve as derived by Houk \& Cowley (\cite{hoco75}).
Pigulski \& Pojma\'{n}ski (\cite{pipo08}) observed it in photometry
as part of the ASAS-3 endeavour and reported the presence of two pulsation
modes with periods 0.22429\,d and 0.45217\,d; they concluded to
a $\beta$ Cep nature for this object. We confirm the variable nature
of this star and the magnitudes measured in the framework of the present
work are given in Table~A2, available at the CDS. These magnitudes
have been analysed using the same multi-frequency Fourier methods
as referred to in the main text. For each filter, we decomposed the
light curve into deterministic processes considering up to three frequencies
and an additional white-noise process. The dominant frequency peaks
are $\nu$~=~2.312, 4.455, 3.349\,d$^{-1}$ for the $B$ bandpass
(semi-amplitudes 0.013, 0.012, and 0.012, respectively), $\nu$~=~1.210,
2.213, and 4.458\,d$^{-1}$ for the $V$ bandpass (semi-amplitudes
0.012, 0.011, and 0.010, respectively), $\nu$~=~4.761, 2.236, and
1.235\,d$^{-1}$ for the $R$ bandpass (semi-amplitudes 0.012, 0.011,
and 0.010, respectively), and, finally, $\nu$~=~1.288, 2.287, and
4.492\,d$^{-1}$ for the $I$ bandpass (semi-amplitudes 0.020, 0.019,
and 0.018, respectively). Considering that the one-day aliasing remains
strong in the sample, we conclude that the dominant frequencies for
the different bandpasses could in some cases share the same single
origin. No frequency other than the dominant frequencies seems to
present some ubiquity among the various filters, indicating that we
are nearing the noise level. We removed the mean magnitude from each
dataset and merged the four light curves aiming at decreasing the
noise. The corresponding Fourier (semi-amplitude) periodogram is given
in Fig.~\ref{fourierHD116827}. Four main frequency peaks can be
spotted at $\nu$~=~2.286\,d$^{-1}$ (semi-amplitude 0.013), $\nu$~=~1.286\,d$^{-1}$
(semi-amplitude 0.011), $\nu$~=~3.319\,d$^{-1}$ (semi-amplitude
0.011), and $\nu$~=~4.355\,d$^{-1}$ (semi-amplitude 0.010). The
one-day alias makes no doubt and we should be careful while making
such a list. We also analysed the global data with the multi-frequency
method introduced by Gosset et al.\ (\cite{Gosset01}). Two independent
frequencies were found to be $\nu$~=~2.286\,d$^{-1}$ (semi-amplitude
0.014) and $\nu$~=~4.493\,d$^{-1}$ (semi-amplitude 0.014), but
no other significant frequency can be spotted. The two derived independent
frequencies are strongly reminiscent of those reported on the basis
of the ASAS-3 photometry ($\nu$~=~4.459\,d$^{-1}$ and $\nu$~=~2.212\,d$^{-1}$),
particularly taking into account the 0.03\,d$^{-1}$ aliasing.

\setcounter{figure}{0} 
\renewcommand{\thefigure}{A\arabic{figure}}
\begin{figure}

\resizebox{9cm}{!}{\includegraphics{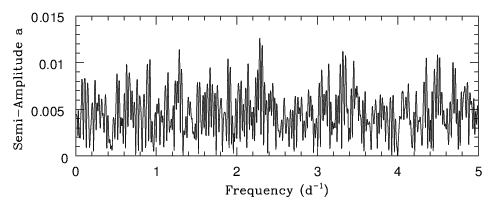}}

\caption{Fourier (semi-amplitude) periodogram for HD 116827.}\label{fourierHD116827}
\end{figure}

\medskip{}

\medskip{}

\noindent \begin{flushleft}
\appendix{\it{A.2. }}{\it{LSS3072 $\equiv$ CPD -61$\degr$3692}}
\par\end{flushleft}

\medskip{}

\noindent This object appears in the catalogue of OB stars of Reed
(\cite{Reed}) and is classified as O9. The magnitudes measured in
the framework of the present work are given in Table~A3, available
at the CDS. It is clearly variable and the light curve is plotted
against time in Fig.~\ref{curveLSS3072}. A search for periods results
in the detection of the possible frequency $\nu$~=~0.1092\,d$^{-1}$
($P$ = 9.16\,d) for each of the four bandpasses and for the combined
data (semi-amplitude 0.023 mag in $B$, 0.023 in $V$, 0.024 in $R$
and 0.023 in $I$). Two bumps are clearly present in Fig.~\ref{curveLSS3072}.
A few discrepant data points are present at HJD\,2\,451\,999.71
and at HJD\,2\,452\,027.54. The deletion of these data points does
not change the result of the period search. However, it is worth noticing
that they are separated by some three entire cycles and could thus
be related to the possible main period. We have no particular reason
to reject these possible outliers. 

\setcounter{figure}{1} 
\renewcommand{\thefigure}{A\arabic{figure}}
\begin{figure}

\resizebox{9cm}{!}{\includegraphics{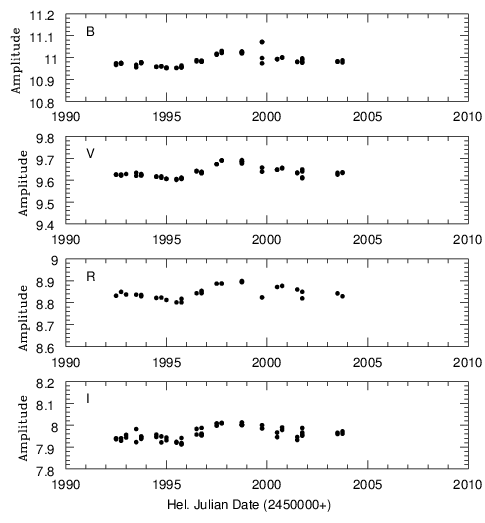}}

\resizebox{9cm}{!}{\includegraphics{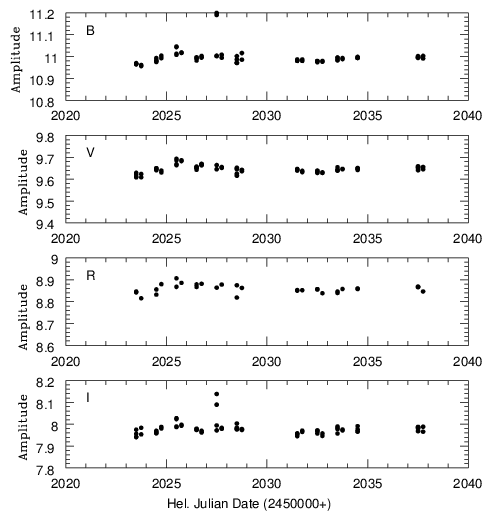}}

\caption{Light curve of LSS 3072.}\label{curveLSS3072}
\end{figure}
\end{document}